\documentclass{mn2e}
\usepackage{epsfig}
\usepackage{longtable}
\usepackage{graphicx}
\usepackage{graphics}
\usepackage{amssymb}
\usepackage{times}
\title[VLT integral field spectroscopy of planetary nebulae]
{Integral field spectroscopy of planetary nebulae: mapping the line diagnostics
and hydrogen-poor zones with VLT FLAMES$^{\ddag}$
\author[Y. G. Tsamis et al.]
{Y. G. Tsamis$^{1, 2}$, J. R. Walsh$^2$, D. P\'{e}quignot$^3$, M. J.
Barlow$^1$, I. J. Danziger$^4$, and X.-W. Liu$^5$\\
$^1$Department of Physics and Astronomy, University College London,
      Gower Street, London WC1E 6BT, UK, e-mail: ygt@star.ucl.ac.uk\\
$^2$Space Telescope European Co-ordinating Facility, European Southern
Observatory, Karl-Schwarzschild Strasse 2, D-85748 Garching, Germany\\
$^3$LUTH, Observatoire de Paris, CNRS, Universit\'e Paris Diderot; 5 Place
Jules Janssen, 92190 Meudon, France.\\
$^4$Osservatorio Astronomico di Trieste, Via G. B. Tiepolo 11, I-34131 Trieste,
Italy.\\
$^5$Department of Astronomy, Peking University, Beijing 100871, PR China\\
\ddag{\it{Based on observations made with ESO telescopes at the Paranal
Observatory under programme ID 075.D-0847(A)}}}}

\date{Accepted 2008 January 28.  Received 2008 January 17; in original form 2007
December 6}
\newcommand{\apj}{ApJ}
\newcommand{\apjs}{ApJS}
\newcommand{\aap}{A\&A}
\newcommand{\aj}{AJ}
\newcommand{\apjl}{ApJL}
\newcommand{\mnras}{MNRAS}
\newcommand{\eld}{$N_{\rm e}$}

\newcommand{\elt}{$T_{\rm e}$}

\newcommand{\exe}{$E_{\rm ex}$}
\newcommand{\cmt}{cm$^{-3}$}

\newcommand{\cpp}{C$^{2+}$}

\newcommand{\opp}{O$^{2+}$}

\newcommand{\foiii}{[O~{\sc iii}]}

\newcommand{\fnii}{[N~{\sc ii}]}

\newcommand{\fariv}{[Ar~{\sc iv}]}

\newcommand{\fneiii}{[Ne~{\sc iii}]}

\newcommand{\nii}{N~{\sc ii}}
\newcommand{\niii}{N~{\sc iii}}

\newcommand{\oii}{O~{\sc ii}}
\newcommand{\cii}{C~{\sc ii}}

\newcommand{\ciii}{C~{\sc iii}}

\newcommand{\hi}{H\,{\sc i}}
\newcommand{\hii}{H~{\sc ii}}
\newcommand{\hei}{He~{\sc i}}
\newcommand{\heii}{He~{\sc ii}}

\newcommand{\hp}{H$^+$}
\newcommand{\hep}{He$^+$}
\newcommand{\hepp}{He$^{2+}$}

\newcommand{\hb}{H$\beta$}
\newcommand{\hg}{H$\gamma$}

\begin{document}
\maketitle

\begin{abstract}

\noindent  Results from the first dedicated study of Galactic planetary nebulae
(PNe) by means of optical integral field spectroscopy with the VLT FLAMES Argus
integral field unit (IFU) are presented. Three typical Galactic-disk PNe have
been mapped with the 11$.''5$ $\times$ 7$.''2$ Argus array: two dimensional
spectral maps of the main shell of NGC\,5882 and of large areas of NGC\,6153
and 7009 with 297 spatial pixels per target were obtained at sub-arcsec
resolutions. A corresponding number of 297 spectra per target were obtained in
the 396.4 -- 507.8\,nm range. Spatially resolved maps of emission lines and of
nebular physical properties such as electron temperatures, densities and ionic
abundances were produced. The abundances of helium and of doubly ionized carbon
and oxygen, relative to hydrogen, were derived from optical recombination lines
(ORLs), while those of \opp\ were also derived from the classic collisionally
excited lines (CELs). The occurrence of the abundance discrepancy problem,
pertaining to oxygen, was investigated by mapping the ratio of ORL/CEL
abundances for \opp\ (the \emph{abundance discrepancy factor}; ADF) across the
face of the PNe. The ADF varies between targets and also with position within
the targets, attaining values of $\sim$ 40 in the case of NGC\,6153 and $\sim$
30 in the case of NGC\,7009. Correlations of the ADF with geometric distance
from the central star and plasma surface brightness (for NGC\,6153), as well as
with \foiii\ electron temperature, plasma ionization state and other physical
properties of the targets are established. Very small values of the temperature
fluctuation parameter in the plane of the sky, $t^2_A$(\opp), are found in all
cases.

It is argued that these results provide further evidence for the existence in
run-of-the-mill PNe of a distinct nebular component consisting of
hydrogen-deficient, super-metal-rich plasma. The zones containing this posited
component appear as undulations in the \cii\ and \oii\ ORL abundance
diagnostics of about 2 spatial pixels across, and so any associated structures
should have physical sizes of less than $\sim$ 1000 astronomical units.
Regarding the origin of the inferred zones, we propose that circumstellar
disks, Abell 30-type knots, or Helix-type cometary globules may be involved.
Implications for emission line studies of nebulae are discussed.

\vspace{0.3cm}


\end{abstract}

\section{Introduction}

Planetary nebulae represent only an ephemeral stage in the late evolution of
low- to intermediate-mass stars, the predominant stellar population in late
type galaxies, but their rich emission line spectra allow us to quantify the
amounts of helium, and heavier elements (mainly carbon, nitrogen, and
$s$-process elements) that these stars produce and return into the interstellar
medium (ISM). Recent studies have indicated that PNe may also contain
non-negligible amounts of endogenous oxygen and neon synthesized by intrinsic
nucleosynthesis in the stellar progenitors and expelled into the nebulae
following the third dredge-up, especially more so in lower metallicity
environments such as those of the Galactic Halo (P\'equignot \& Tsamis 2005)
and of satellites of the Galaxy (P\'equignot et al. 2000; Zijlstra et al. 2006;
Leisy \& Dennefeld 2006). Their importance as vital sources of neutron-capture
elements, such as for instance selenium and krypton among others, is now
observationally confirmed (P\'equignot \& Baluteau 1994; Sterling et al. 2007).
With the advent of 8-m class telescopes, PNe are increasingly being used as
probes of the chemical evolution of Local Group galaxies (e.g., Gon{\c c}alves
et al. 2007; Saviane et al. 2008), while by virtue of their intrinsically
bright monochromatic \foiii\ 500.7\,nm emission they have been detected in the
intracluster stellar population of the Coma cluster at a distance of 100\,Mpc
(Gerhard et al. 2005, 2007).

A focal problem, however, remains in the astrophysics of these objects: namely
that large discrepancies are measured between abundances and temperatures
obtained from recombination lines and continua {\sl versus} those obtained from
the bright forbidden lines (e.g., Tsamis et al. 2003b, 2004; Liu et al. 2000,
2004; Wesson et al. 2005). For the majority of more than 100 PNe surveyed thus
far by long-slit (or \'{e}chelle) spectroscopy, abundances of the elements
carbon, oxygen, nitrogen and neon derived from their optical recombination
lines (ORLs) are higher than those derived from the classic (`forbidden')
collisionally excited lines (CELs) by factors 2--3 (see review by Liu 2006 and
references therein). For about 5--10 per cent of the surveyed objects however
the discrepancies are in the 4--80 range, with the most pathological case to
date being Hf\,2-2 (Liu et al. 2006), a southern planetary nebula. In the cases
of carbon, nitrogen (Tsamis 2002; Tsamis et al. 2004) and oxygen (Liu et al
2001; Tsamis et al. 2004), large metal overabundances are correlated with lower
plasma temperatures derived from ORLs and continua than from forbidden lines.
The resolution of this problem remains of high priority as it continues to
undermine our trust on secure elemental abundance determinations for PNe, and
for other types of emission line objects (such as \hii\ regions; e.g. Tsamis et
al 2003a).

In the overwhelming majority of cases the ratio of ORL/CEL abundances for a
given heavy element ion (the \emph{abundance discrepancy factor}, ADF) is not
correlated with the excitation energies, \exe, of the CELs involved: meaning
that, per individual PN, infrared (IR) or ultraviolet (UV)/optical CELs of low
and high \exe\ respectively, yield very comparable ionic abundances, which are
uniformly lower than the corresponding abundances of the same ions derived from
ORLs. The most likely explanation for these spectroscopic results was judged to
be the presence of cold plasma regions embedded in the nebular gas in the form
of relatively dense, hydrogen-poor condensations -- clumps or filaments (Liu et
al. 2000; Tsamis 2002, Tsamis et al. 2004). Due to their elevated content in
heavy elements (several times Solar), these plasma regions would have reached
photoionization equilibrium at lower temperatures than the ambient `normal'
composition gas by emitting far-IR lines, the primary nebular thermostat in
relatively low temperatures. Since the emissivity of heavy element
recombination lines is enhanced at lower temperatures, while at the same time
that of the classic forbidden lines is diminished, the hydrogen-poor, heavy
element-rich clumps would emit heavy element recombination lines profusely,
yielding a truer estimate of the heavy element content of these regions.

Whereas the long-slit surveys yielded concrete evidence for elevated
recombination-line abundances, the temperature of the suspected
hydrogen-deficient clumps was tougher to determine since the diagnostic lines
involved are faint and often suffer from blends in lower resolution spectra.
Nevertheless, there has been evidence from those studies for temperatures lower
by several thousand K than the typical CEL temperatures of photoionized nebulae
(Tsamis et al 2004; Liu et al. 2004). This is one aspect of the proposed `dual
abundance model' solution that renders it self-consistent, something rather
lacking from alternative propositions, such as small-scale temperature
fluctuations in a chemically homogeneous medium (e.g. Peimbert et al. 2004),
which do not invoke a physical process to explain the discrepancies. A major
unresolved issue remains the as yet unknown origin of the high abundance
clumps, and their mass relative to the total ionized nebular gas.

Heavy element recombination lines in nebulae can be a thousand times or more
fainter than hydrogen recombination lines, in sharp contrast to the luminous
forbidden lines emitted by the same heavy ions. We therefore brought to bear on
this thorny issue the combination of integral field spectroscopy (IFS) and the
light collecting power of the Very Large Telescope (VLT). IFS techniques allow
spectra to be obtained from every spatial element in a two-dimensional field of
view, combined with imaging capability of the field at any wavelength across a
given $\lambda$ range. This is usually achieved by means of an IFU array
composed of microlenses coupled to optical fibres or slicers (e.g.
Allington-Smith 2006). These are fed into a spectrograph and are subsequently
individually imaged on the focal plane of the telescope.

The contiguous array of microlenses (or spatial pixels -- `spaxels') can also
be thought of as a series of long-slits stacked together and, in this sense, an
IFU of $m$ $\times$ $n$ spaxels used on a single telescope is equivalent to
using an array of $m$ telescopes equipped with long slits of $n$ pixels each.
It could be argued that in the former case a given observing programme is
executed $m$ times faster than when using a classic long-slit spectrograph. In
reality this is not strictly true as invariably there are some light losses
associated with IFU fibres and their optical connection to the spectrograph
(particularly at the interfaces with the microlenses and the feed into the
slit). Overall, IFU observations afford considerable advantages especially in a
science programme like ours where faint diagnostic lines are targeted, and
where 2D mapping of nebular regions is sought in the hope of isolating suspect
metal-rich regions which may {\it a priori} have a random spatial distribution
in the nebula.

In addition, IFS data can provide the much-needed input to fully realize the
capabilities of new plasma codes such as those developed in recent years for
the 3D modelling of photoionized nebulae (e.g. Ercolano et al. 2003; Morisset
et al. 2005). For example, such codes can produce simulated 3D or 2D spectral
images of model nebulae at arbitrary wavelengths and inclinations that can be
readily evaluated using images obtained by IFS. Moreover, high spectral
resolution IFS data such as those we obtained for NGC\,7009 (Walsh et al. in
preparation) can yield detailed information on the velocity field of an object
that could allow a 3D kinematical model to be built. These are some of the
wider aspects of this work.

Here we pursue a physicochemical analysis of three representative PNe belonging
to the Galactic-disk population with published oxygen ADFs ranging from $\sim$
2 -- 10. Long-slit spectroscopy of the targets has been published: NGC\,5882
(Tsamis et al. 2003b, 2004); NGC\,6153 (Liu et al. 2000); NGC\,7009 (Liu et al.
1995; hereafter LSBC). In those studies, the first two targets were observed
with a long slit scanned across the whole nebular surface. This technique,
which was applied throughout the long-slit surveys, yielded \emph{mean}
relative emission line fluxes for the whole PNe, but naturally any variations
of the physical properties in the direction perpendicular to the slit could not
be sampled. An analysis along a NGC\,6153 fixed slit sampling a `slice' of the
nebula revealed substantial variations of the physical properties over the slit
length. Typically, however, the formerly employed $\sim$ 2~arcsec-wide slits
achieved a rather coarse spatial sampling of $\sim$1.6~arcsec per pixel,
without sampling the seeing which was of the order of 1~arcsec.

We obtained contiguous 2D spatial coverage of these PNe at sub-arcsec
resolution with the Fibre Large Array Multi Element Spectrograph
(FLAMES)/Giraffe IFU Argus. Our main aim was to map out the nebular properties
and their variations across large nebular areas and to investigate in detail
the existence of any heavy element-rich nebular regions, so as to place
constraints on models that have been put forward to explain this astrophysical
quandary. The sample PNe were chosen because NGC\,6153 and NGC\,7009 exhibit
the abundance anomaly problem to a high degree (with known oxygen ADFs of
$\sim$ 10 and 5 respectively) and are excellent cases for a detailed study,
whereas NGC\,5882 (with an ADF of $\sim$ 2) was meant to be our `control'
target as it represents the majority of Galactic nebulae that show a moderate
degree of abundance discrepancies. The paper layout is as follows: the
specifications of the observational data set are discussed in Section~2, the
data reduction methods are presented in Section~3, and the results in Sections
~4 and 5. The discussion and our conclusions appear in Section~6. An
introduction to this study was presented by Tsamis et al. (2007).


\section{Observations}

\setcounter{figure}{0}
\begin{figure}
\centering \epsfig{file=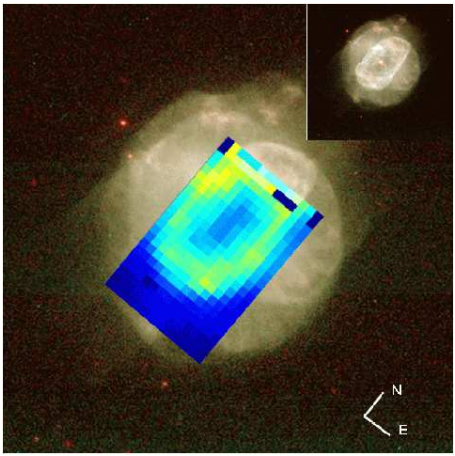, width=6. cm, scale=, clip=, angle=0}
\caption{NGC\,5882: VLT Argus \hb\ map overlaid on a {\it HST} WFPC2 image of
the nebula covered by the $11.''5$ $\times$ $7.''2$ IFU. The {\it HST} snapshot
was taken with a broad-band F555W filter (credit: H. Bond and Space Telescope
Science Institute). The position of the stellar nucleus is visible in the
inset.}
\end{figure}

\setcounter{figure}{1}
\begin{figure}
\centering \epsfig{file=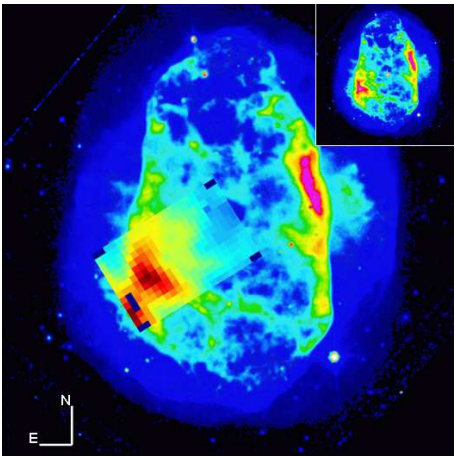, width=6. cm, scale=, clip=, angle=0}
\caption{NGC\,6153: VLT Argus \hb\ map overlaid on a pseudocolour \emph{HST}
WFPC2 broad-band filter F814W image of the nebula covered by the $11.''5$
$\times$ $7.''2$ IFU (source of Hubble image: Liu et al. 2000). The position of
the stellar nucleus is visible in the inset.}
\end{figure}

Integral field spectra of NGC\,5882, 6153 and 7009 were obtained in service
mode between March--June 2005 in sub-arcsec seeing with the FLAMES/Giraffe
Argus IFU on the 8.2-m VLT UT2/Kueyen (P.I. Tsamis; ESO Period 75). The
observing log is presented in Table~1. All targets were observed in the 396.4
-- 507.8\,nm range with the large $11.''5$ $\times$ $7.''2$ IFU. The Argus unit
is a rectangular array of 22 $\times$ 14 microlenses fed by optical fibres: the
$11.''5$ $\times$ $7.''2$ array thus projects 308 spaxels on the sky; the
individual spaxel size is $0.''52$ $\times$ $0.''52$. Eight spaxels are
reserved for sky subtraction purposes (from a total of 14 dedicated sky-fibres)
and three currently correspond to broken fibres (appearing in the second row of
a reconstructed Argus image; see Fig.\,~6); the remainder provide 297
individual spectra per target. The Giraffe spectrograph is equipped with a 2048
$\times$ 4096 EEV CCD which has 15\,$\mu$m pixels. Two separate wavelength
ranges were observed using the LR02 (3964--4567\,\AA) and LR03
(4501--5078\,\AA) grating set-ups. The respective spectral resolving powers,
$R$ ($=$ $\lambda$/$\Delta\lambda$), were 10\,200 and 12\,000. Higher spectral
resolution HR04 and HR06 spectra ($R$ $=$ 32\,500) of NGC\,7009 were also taken
with the small $6.''6$ $\times$ $4.''2$ IFU and will be published in a separate
paper (Walsh et al. in preparation). Total exposure times were 6930\,sec
(NGC\,5882), 24\,000\,sec (NGC\,6153), and 4260\,sec (NGC\,7009); more time was
devoted to NGC\,6153 than to the other two targets since this nebula has the
lowest surface brightness, and exhibits the abundance anomaly to a high degree.
The average seeing during the NGC\,5882 exposures was 0.6 arcsec full width at
half maximum (FWHM); during six out of ten observing blocks of NGC\,6153 it was
better than 0.8 arcsec, while during the remaining four it was $\sim$ 0.9
arcsec. During the NGC\,7009 exposures the seeing ranged between 0.5--0.8
arcsec. The sky transparency conditions were generally in the `clear' category.

The positions observed on each PN were given due consideration. In Figs.\,~1--3
we show the exact positions observed by overlaying Argus spectral images taken
in the light of \hb\ 4861\,\AA\ on \emph{Hubble Space Telescope} ({\it HST})
WFPC2 images of the nebulae. The Argus maps have been drawn to scale in all
cases. Regarding NGC\,5882 (Fig.\,~1), which is the nebula with the smallest
apparent size of the three, the IFU array was positioned so that the largest
part of the bright shell of the PN as well as part of the fainter halo to the
south could be covered. The centre of the IFU was positioned at $\alpha$ $=$
15$^{\rm h}$\,16$^{\rm m}$\,49$^{\rm s}$.8, $\delta$ $=$ $-$45$^{\rm
o}$\,39$'$\,01$''$.4 (J2000).

For NGC\,6153 (Fig.\,~2) we chose to align the large axis of Argus roughly
along the minor axis of the PN so as to cover both the bright patch visible on
the {\it HST} image in the southeastern outskirts of the nebula and the inner
fainter regions closer to the central star -- this position also shares the
orientation of the 2-arcsec wide fixed long slit employed by Liu et al. (2000).
The centre of the IFU was positioned at $\alpha$ $=$ 16$^{\rm h}$\,31$^{\rm
m}$\,31$^{\rm s}$.0, $\delta$ $=$ $-$40$^{\rm o}$\,15$'$\,12$''$.4 (J2000).

NGC\,7009 (Fig.\,~3) has an elliptical shape with low ionization emission
regions at both ends of its major axis of symmetry which give it a bipolar
appearance. It shows a clearly defined bright inner shell surrounded by a
fainter outer envelope. We thus chose to place the IFU array so that both the
faint regions in the centre of the PN and the sharp rim of the bright inner
shell could be covered. The centre of the IFU was positioned at $\alpha$ $=$
21$^{\rm h}$\,04$^{\rm m}$\,10$^{\rm s}$.8, $\delta$ $=$ $-$11$^{\rm
o}$\,21$'$\,48$''$.7 (J2000). In all cases the central stars of the PNe
(visible in the insets of the {\it HST}/Argus overlays) fell within the IFU
field of view and were a useful point of reference. This further allowed us to
examine the nebular physical conditions in the immediate vicinity of the
stellar nuclei.

\setcounter{table}{0}
\begin{table}
\centering
\begin{minipage}{80mm}
\caption{Journal of VLT 8.2-m FLAMES Argus observations.}
\begin{tabular}{lccccc}
\noalign{\vskip3pt}\noalign{\hrule}\noalign{\vskip3pt}
Target             &Date                     &$\lambda$-range &Grating &Exp. time   \\
               &(UT)                     &(\AA)           &        &(sec)               \\
\noalign{\vskip3pt}\noalign{\hrule}\noalign{\vskip3pt}
NGC\,5882      &23/03/05                 &3964--4567      &LR02          &3$\times$440      \\
               &23/03/05                 &4501--5078      &LR03          &3$\times$330      \\
               &29/04/05                 &3964--4567      &LR02          &6$\times$440      \\
               &29/04/05                 &4501--5078      &LR03          &6$\times$330    \\
NGC\,6153      &29/04/05                 &3964--4567      &LR02          & 6$\times$800         \\
               &29/04/05                 &4501--5078      &LR03          & 6$\times$400         \\
               &30/04/05                 &3964--4567      &LR02          & 12$\times$800        \\
               &30/04/05                 &4501--5078      &LR03          & 12$\times$400        \\
               &01/05/05                 &3964--4567      &LR02          &  4$\times$800        \\
               &01/05/05                 &4501--5078      &LR03          &  4$\times$400        \\
NGC\,7009      &02/06/05                 &3964--4567      &LR02          &  8$\times$270        \\
               &02/06/05                 &4501--5078      &LR03          & 12$\times$175        \\
\noalign{\vskip3pt}\noalign{\hrule}\noalign{\vskip3pt}
\end{tabular}
\end{minipage}
\end{table}

\setcounter{figure}{2}
\begin{figure}
\centering \epsfig{file=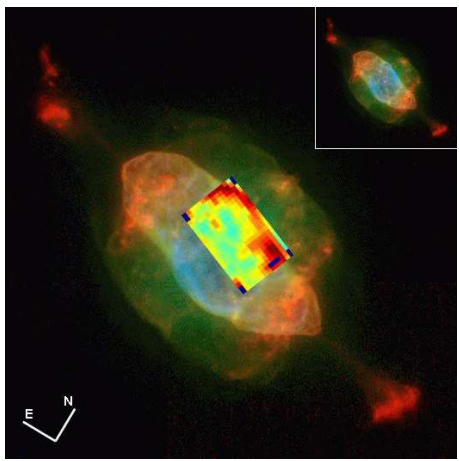, width=6. cm, scale=, clip=, angle=0}
\caption{NGC\,7009: VLT Argus \hb\ map overlaid on a {\it HST} WFPC2 image
outlining the area of the nebula covered by the $11.''5$ $\times$ $7.''2$ IFU.
The {\it HST} pseudocolour is a composite of F658N $=$ \fnii\ (red), F502N $=$
\foiii\ (green) and F469N $=$ \heii\ (blue) frames (credit: B. Balick and Space
Telescope Science Institute). The position of the stellar nucleus is seen in
the inset.}
\end{figure}

\section{Data reduction}

The spectra were reduced at ESO Garching with the girBLDRS
pipeline\footnote{The girBLDRS pipeline is provided by the Geneva Observatory
(http://girbldrs.sourceforge.net/)} in a process which included bias removal,
localization of fibres on the flats, extraction of individual fibres,
wavelength calibration and rebinning of the ThAr lamp exposures, and the full
processing of the science frames which resulted in cosmic-ray cleaned,
flat-fielded, wavelength-rebinned spectra. In practice the pipeline actions
were initiated by the creation of a master bias frame from the raw biases ({\sc
biasMast} pipe) which was subtracted from the science exposures at a later
reduction step. The localization of fibres on the flat field frames followed
({\sc locMast} pipe). The full master wavelength solution was produced via the
extraction, wavelength calibration and rebinning of the raw Th-Ar exposures
({\sc wcalMast} pipe). This was followed by the full processing of the science
frames via the {\sc extract} recipe which, by using the products of the
previous steps, subtracts the master bias from the science exposures, takes out
an average rejecting any cosmic ray hits, performs the flat-fielding, and
finally rebins the extracted spectra in wavelength space. In this way, four to
six science frames per observing block were processed together and the
resulting frame was corrected for atmospheric extinction within {\sc iraf}. The
airmass during the exposures was $\sim$ 1.1. When this varied by more than
$\sim$ 0.1 during the course of a single observing block (1 hr) the frames were
extracted individually and were corrected for atmospheric extinction one at a
time. They were subsequently averaged using the {\sc imcombine} task which
performed the cosmic ray rejection as well. The flux calibration was achieved
using exposures of the white dwarf EG~274 using the tasks {\sc calibrate} and
{\sc standard}. The flux standard exposures were first individually extracted
with girBLDRS and the spaxels containing the star were summed up to form a 1D
spectrum. The sensitivity function was determined using {\sc iraf}'s {\sc
sensfunc} and this was subsequently applied to the combined science exposures.
The sky subtraction was performed by averaging the spectra recorded by the sky
fibres and subtracting this spectrum from that of each spaxel in the IFU. Each
of the 14 sky-fibre spectra were examined and the average count in each
compared. Four sky fibres had higher counts caused by contamination from the
simultaneous calibration lamp (see below), and were not included in the
computation of the mean sky spectrum.

\setcounter{figure}{3}
\begin{figure}
\centering \epsfig{file=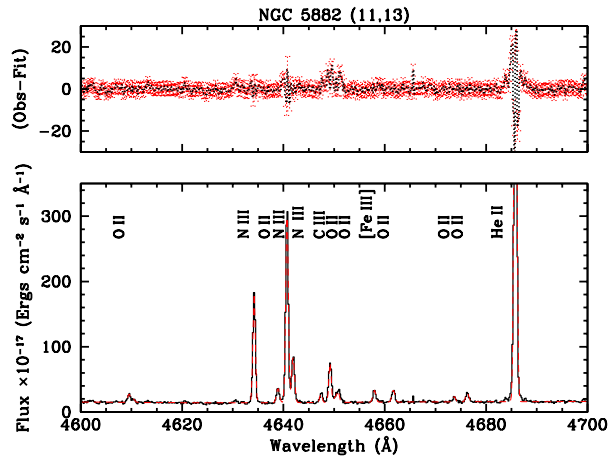, width=6. cm, scale=, clip=, angle=-90}
\caption{Spectrum of NGC\,5882 from spaxel (11, 13) highlighting the \oii\ V1
multiplet recombination lines. The red line is a sum of Gaussians plus a cubic
spline continuum fit to the data (black line). The top panel shows the
residuals of the fit (black) compared to the error bars of each point in the
spectrum (red).}
\end{figure}

\setcounter{figure}{4}
\begin{figure}
\centering

\epsfig{file=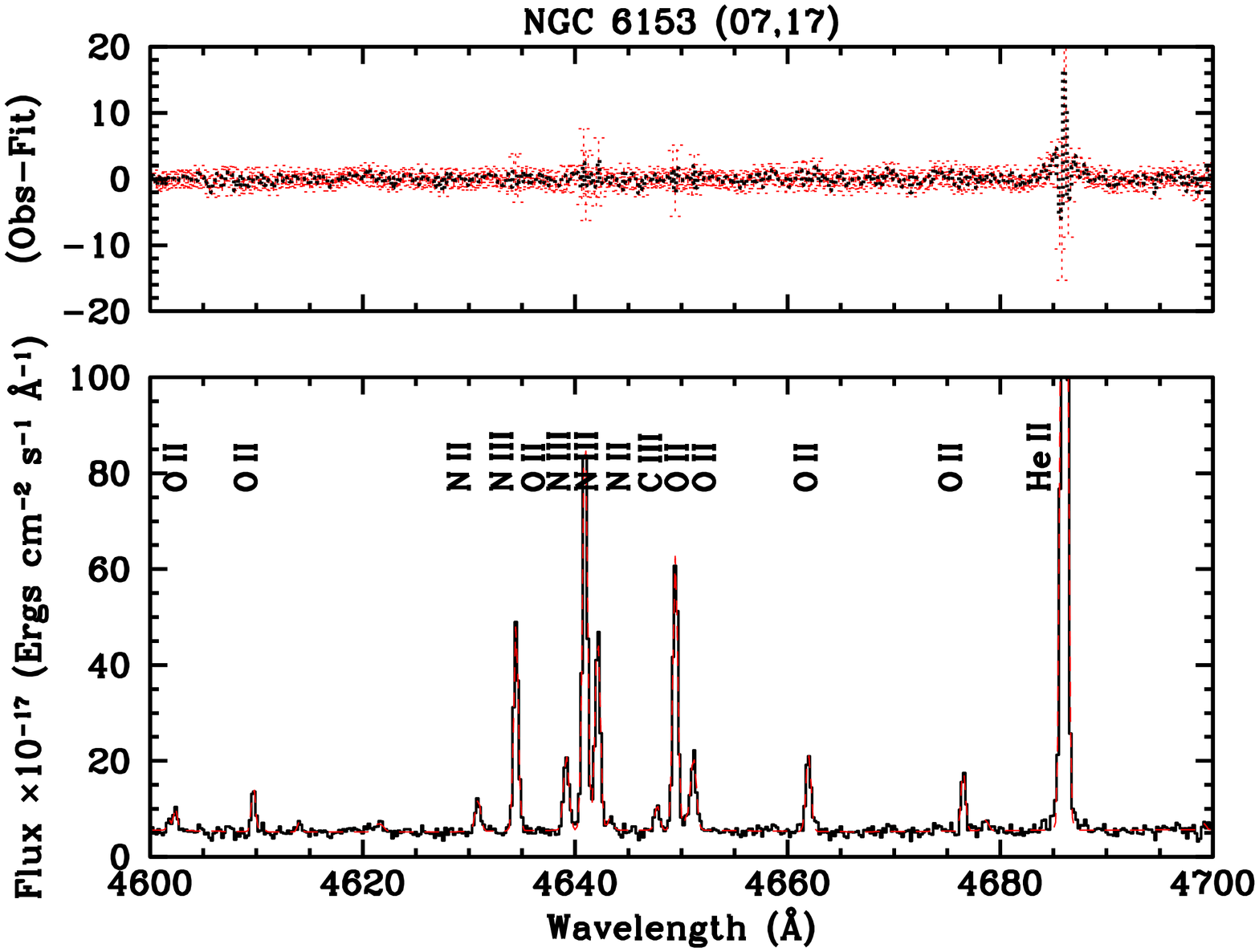, width=6. cm, scale=, clip=, angle=-90}
\epsfig{file=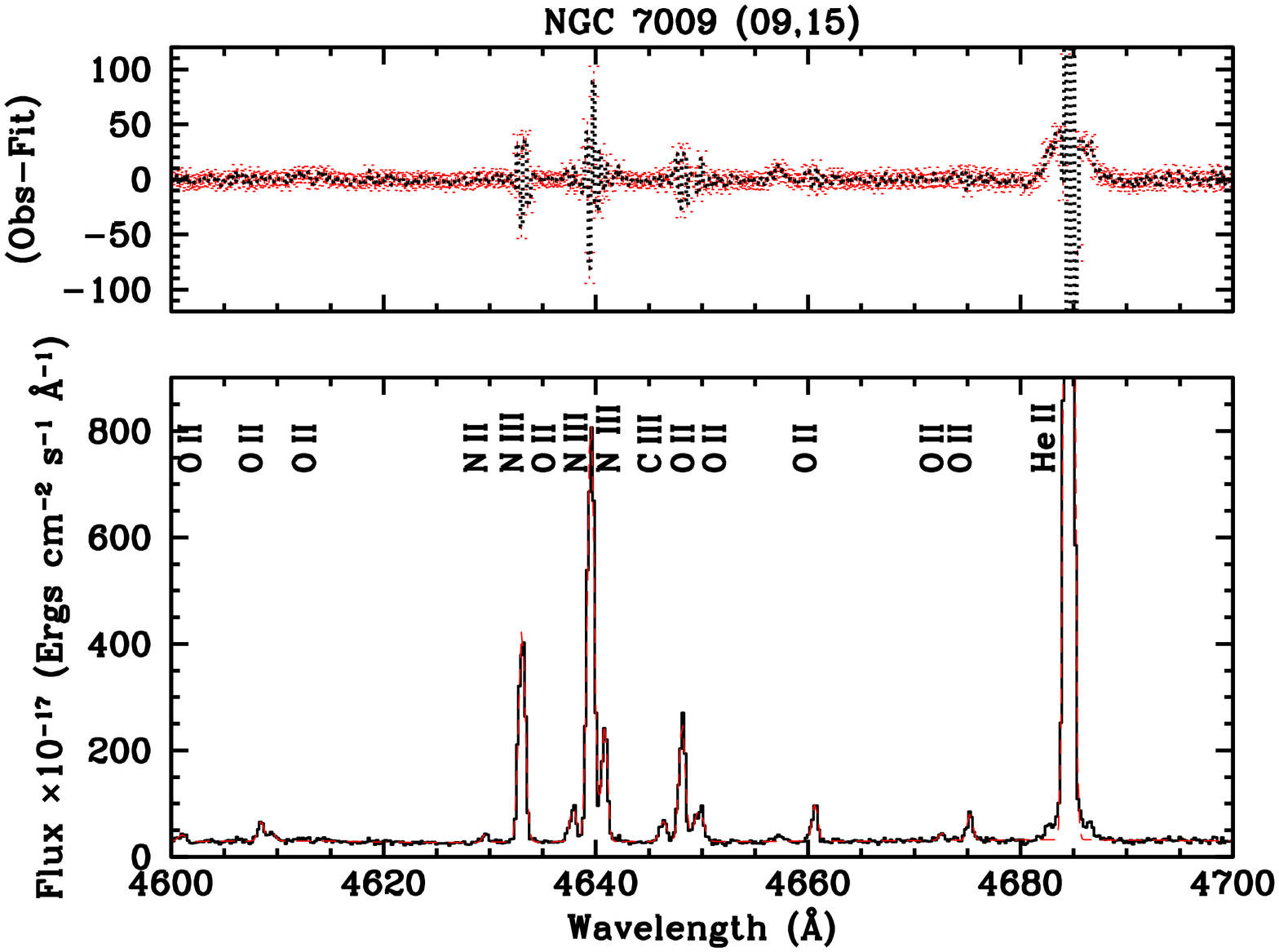, width=6. cm, scale=, clip=, angle=-90}

\caption{Same as Fig.\,~4 for NGC\,6153 (top) and NGC\,7009 (bottom).}

\end{figure}

Custom-made scripts allowed us to convert the row by row stacked, processed CCD
spectra to data cubes. The nebular emission line fluxes were generally measured
automatically by fitting Gaussian profiles via means of the dedicated program
{\sc spectre} which is based on {\sc minuit} $\chi^2$ minimization routines
(James 1998). This enabled us to retrieve from the spectra associated with each
spaxel: the line fluxes, the line FWHM, equivalent widths (EW), line centroids,
and the respective errors of these quantities which were propagated in all
physical properties derived afterwards. \footnote{The pre-analysis errors
produced by the pipeline were used as the starting point of the error analysis;
they also include uncertainties from the flat-fielding and the geometric
correction of the spectra and not just photon-noise (Blecha \& Simond 2004).}
The same program was used to interactively fit the nebular lines in the
vicinity of the PN nuclei, and the absorption lines underneath the \hi\ lines
over the same spaxels. In Figs.\,~4 and 5 Giraffe spectra of the PNe from
single spaxel extractions in the vicinity of 465.0 nm are shown, covering the
full \oii\ V1 recombination multiplet, a few \nii\ and \ciii\ ORLs, \niii\
Bowen fluorescence lines, and \heii\ $\lambda$4686. The lower panels display
the observed spectrum (black line), along with the sum of the fitted continuum
(cubic spline fit) and the individual Gaussians (red line). The upper panels
show the errors bars at each point in the spectrum (red) and the residuals from
the line and continuum fitting (black). In general the fit residuals are
comparable to the error bars. The fits in Figures 4--5 are for single Gaussian
line fits to each emission species; for NGC\,6153, 130 spaxels were fitted with
double Gaussians, since the velocity splitting (due to nebular expansion) of
0.6\,\AA\ could be resolved by two lines of FWHM 0.5\,\AA. In the case of
NGC\,5882, 30 spaxels were fitted by double profiles.

The Gaussian fits to the brightest lines in Figures 4--5 show that the detailed
profile deviates from a Gaussian and in the case of NGC\,7009, for example, the
very strong \heii\ $\lambda$4686 line shows residuals in the line wings that
are greater than the errors. Since only Gaussians were used for fitting,
without a detailed line profile describing the grating scattering line wings,
these line wing features were not fitted in detail. Neglect of these faint
features constitutes a loss of less than one per cent to the line flux, and it
was considered that a consistent line profile shape for all lines (i.e.
Gaussian) was of highest priority for comparison of relative line fluxes (on
which plasma conditions and abundance determinations that are central to this
analysis depend). In general, the spectral resolution achieved was sufficient
to resolve any known blends that affect heavy element ORLs in lower resolution
spectra. Extra care was taken to interactively fit any remaining blended lines
and to also fit double Gaussian profiles in cases where the spectral resolution
was high enough to resolve the expansion of the nebula (e.g. in the case of
NGC\,6153).

\setcounter{figure}{5}
\begin{figure}
\centering \epsfig{file=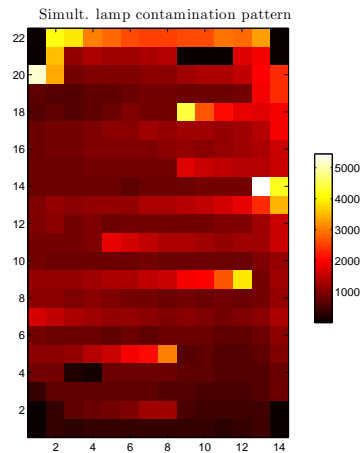, width=9. cm, scale=, clip=, angle=0}
\caption{A spectral map of the continuum showing the characteristic pattern
caused by contamination of the CCD from light leaking from the simultaneous
calibration lamp of FLAMES. See text for details.}
\end{figure}

An unforeseen issue arose due to the use of the `simultaneous calibration lamp'
during the science exposures. Use of this device is not really necessary for
spectrophotometric work and it was left on unintentionally during the
preparation of phase II. The lamp introduced an additive contamination pattern
(Francesca Primas, private communication) on certain spaxels of the array (see
Fig.\,~6), which showed up as a mix of broad and narrow features. All these
features are light scattered from fibre-to-fibre and the broad features are
presumably the scattering wings of strong lines in the comparison lamp
spectrum. Since the PN line emission covered the whole of the IFU area for all
three nebulae, there was no opportunity with the current data to subtract PN
emission-free spaxel spectra in order to remove the contamination pattern.
Instead the contaminated spectra were interactively fitted by following the
continuum in detail and any spurious narrower `lines' were fitted by additional
Gaussians. The results were satisfactory, although of course the fit errors on
the nebular line fluxes were increased at those spaxels. The spectra of
NGC\,6153, which is the nebula with the lowest surface brightness, were most
affected and 40 spaxels were interactively fitted and decontaminated in the
manner described above.

Likewise, a scattered light feature appeared towards the last few CCD columns
(long wavelengths) at the top-right corner of the stacked array. This was due
to a residual CCD signal, which is a known cosmetic defect of the chip. The
signal is additive (in excess of the bias) and, as it affected only certain
spaxels in the reconstructed Argus image, it was fitted with a low-order
polynomial and subtracted out. Finally, apart from the three fibres which were
broken during ESO Period 75, the fibre corresponding to spaxel (4,
4)\footnote{(X, Y) spaxel coordinates in the text are measured respectively on
the minor and major axis of the Argus IFU; see Fig\,~6.} of the reconstructed
array is problematic in that it exhibits an anomalous spike in its signal,
which often affected spaxels (3, 4) and (5, 4) too -- the respective spectra
were therefore handled accordingly and generally suffer from large
uncertainties.

\section{Results}

After the line fluxes had been measured, the reconstructed Argus data cubes
were corrected for interstellar extinction using the $c$(\hb) extinction maps
derived from a comparison of the observed to the predicted \hg/\hb\ line ratio
using its case B value from Storey \& Hummer (1995). The Galactic extinction
law of Howarth (1983) with a ratio of total to selective extinction ($R_V$) of
3.1 was used. Since the LR02 and LR03 spectra, containing the H$\gamma$ and
H$\beta$ lines respectively, were taken in different exposures, and often on
different nights, then it was necessary to tie the fluxes together in order to
determine the extinction of the maps from the H$\gamma$ to H$\beta$ ratio. This
comparison was performed by taking the total flux from the H$\gamma$ and
H$\beta$ maps and comparing the ratio with that observed from long-slit
studies. For NGC\,5882 the absolute value of the ratio was taken as 0.410 from
the scanned-slit (mean over the whole PN) data of Tsamis et al. (2003b); for
NGC\,6153, the ratio for the whole nebula (0.333) was taken from Liu et al.
(2000); for NGC\,7009, the ratio of 0.443 for a value of $c$(\hb) $=$ 0.20 was
adopted (cf. LSBC). It should be noted that the IFU area does not exactly
correspond to the comparison regions from the long-slit data, so the absolute
extinction, and hence dereddened line ratios can be systematically offset. To
gain an idea of the photometric quality of the observations, the LR02 and LR03
maps had to be corrected by 47, 5 and 5 per cent for NGC\,5882, 6153 and 7009
respectively.

Potential shifts between the LR02 and LR03 maps, caused either by a slightly
different pointing between the observations (although note that the LR02 and LR03
setting were always taken in pairs; Table~1) and/or the changing differential
atmospheric extinction between the settings, were determined by
cross-correlation of features in the H$\gamma$ and H$\beta$ maps. For NGC\,5882
a shift of 0.25 spaxels in the IFU X-direction could be clearly detected (this
PN has the sharpest spatial features as it was observed under $\sim$ 0.6 arcsec
seeing). It was nonetheless decided not to correct the maps for this small
shift since data at the IFU edges would be compromised. The extinction map for
NGC\,5882 does not show a very pronounced cusp at the position of the shell, as
might be expected if the H$\gamma$ and H$\beta$ maps were poorly aligned.

In the paragraphs that follow we present spectral maps for representative
emission lines, the plasma physical conditions which were based on spectral map
ratios, and the resulting chemical abundances.

\setcounter{figure}{6}
\begin{figure*}

\centering \epsfig{file=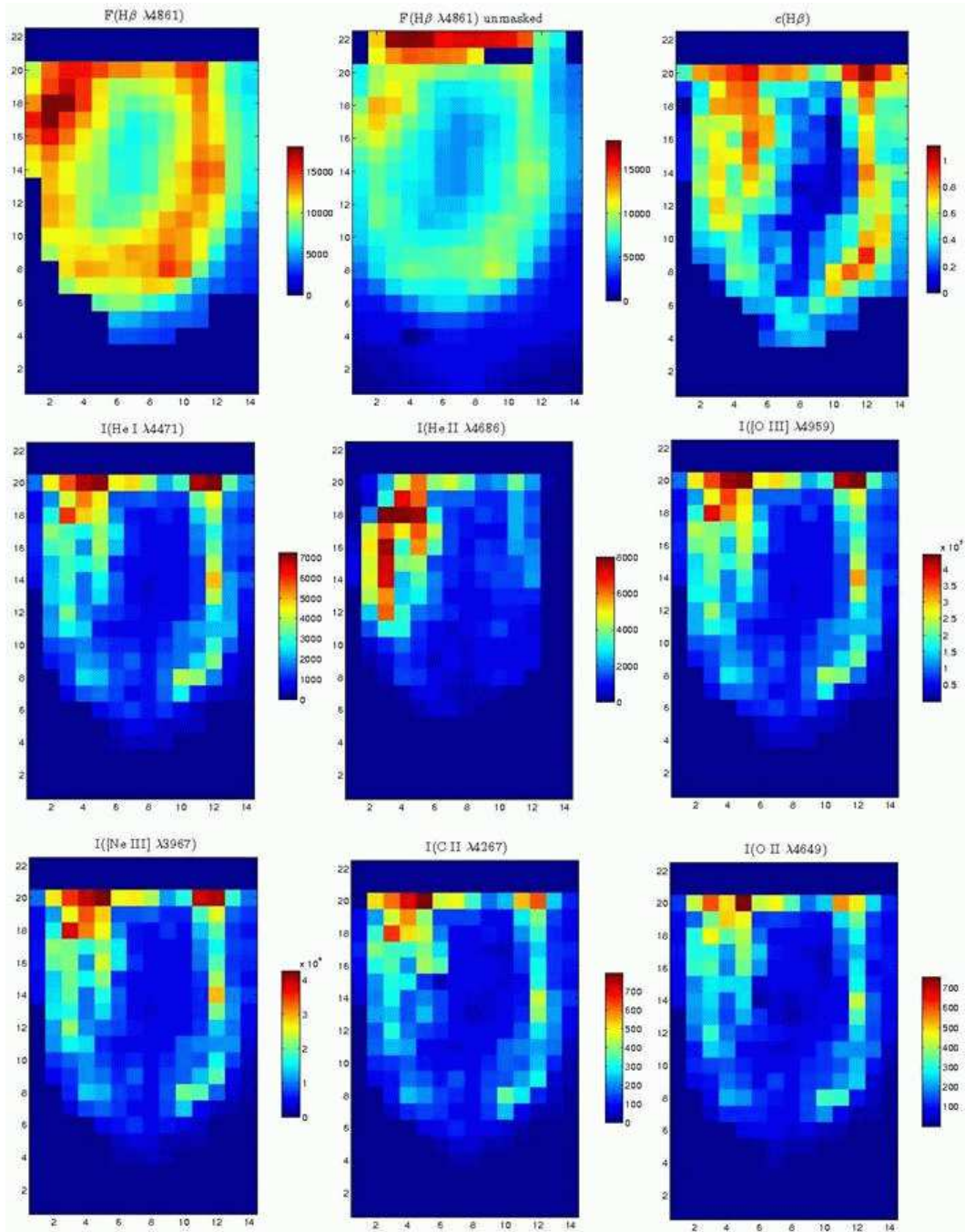, width=14. cm, scale=, clip=, angle=0}

\caption{NGC\,5882 monochromatic line flux ($F$) and dereddened intensity ($I$)
maps (in units of 10$^{-17}$ erg\,cm$^{-2}$\,s$^{-1}$), along with the
extinction constant employed, $c$(\hb). Blank spaxels: three in 2nd row from
top (broken Argus fibres); two per corner (fibres reserved for sky
subtraction). The top two rows have not been considered in the physical
analysis -- see text for details. The central star position is at coordinates
(6, 15). North is up and east is to the right (cf. Fig.\,~1).}

\end{figure*}

\setcounter{figure}{7}
\begin{figure*}
\centering

\epsfig{file=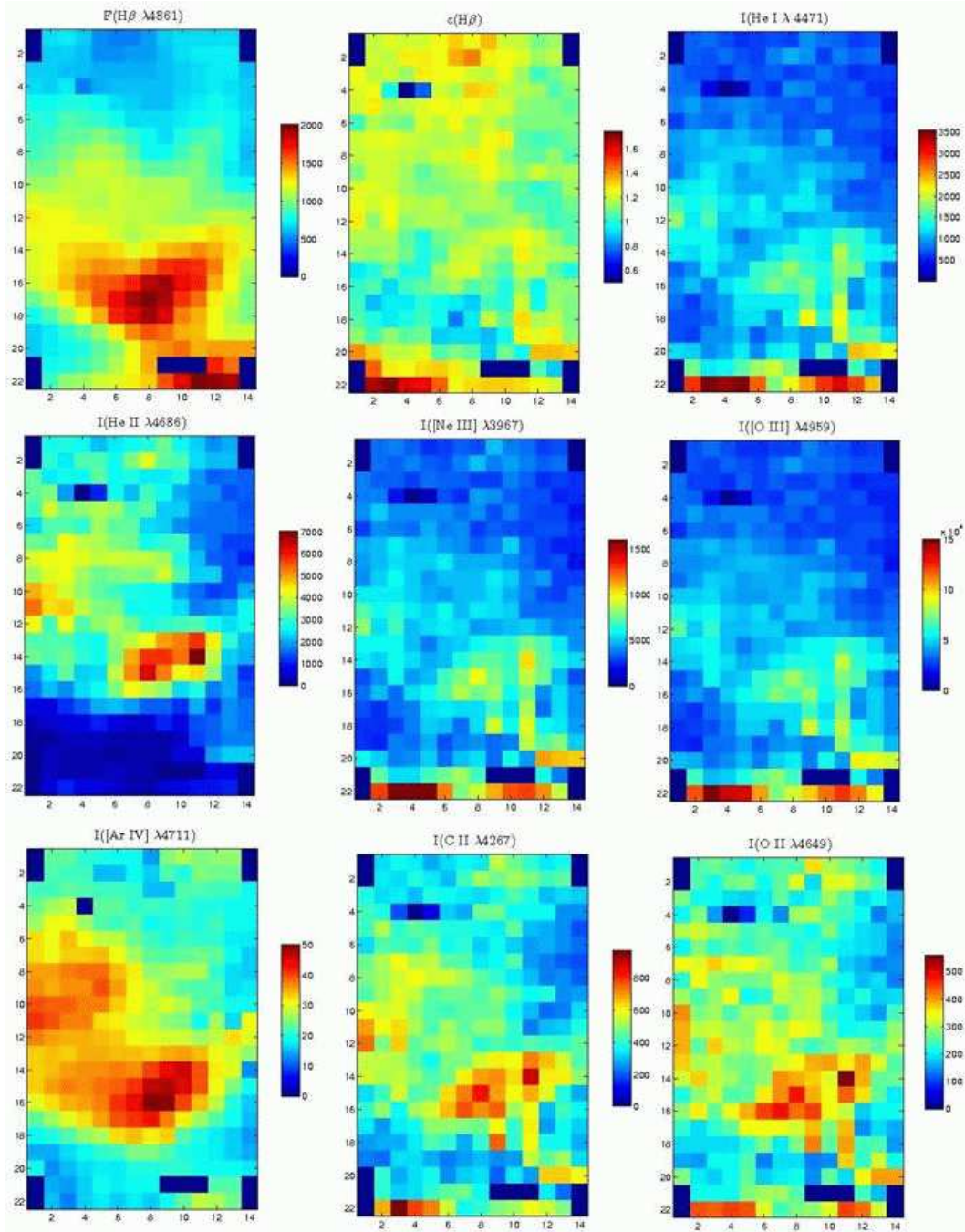, width=14. cm, scale=, clip=, angle=0}

\caption{NGC\,6153 monochromatic line flux ($F$) and dereddened intensity ($I$)
maps (in units of 10$^{-17}$ erg\,cm$^{-2}$\,s$^{-1}$), along with the
extinction constant employed, $c$(\hb). Blank spaxels: three in 2nd row from
bottom (broken Argus fibres); two per corner (fibres reserved for sky
subtraction). In this and subsequent figures spaxels (3--5, 4) suffer from
large uncertainties. The central star position is at coordinates (9, 3).
North-west is up (cf. Fig.\,~2).}

\end{figure*}

\setcounter{figure}{8}
\begin{figure*}
\centering

\epsfig{file=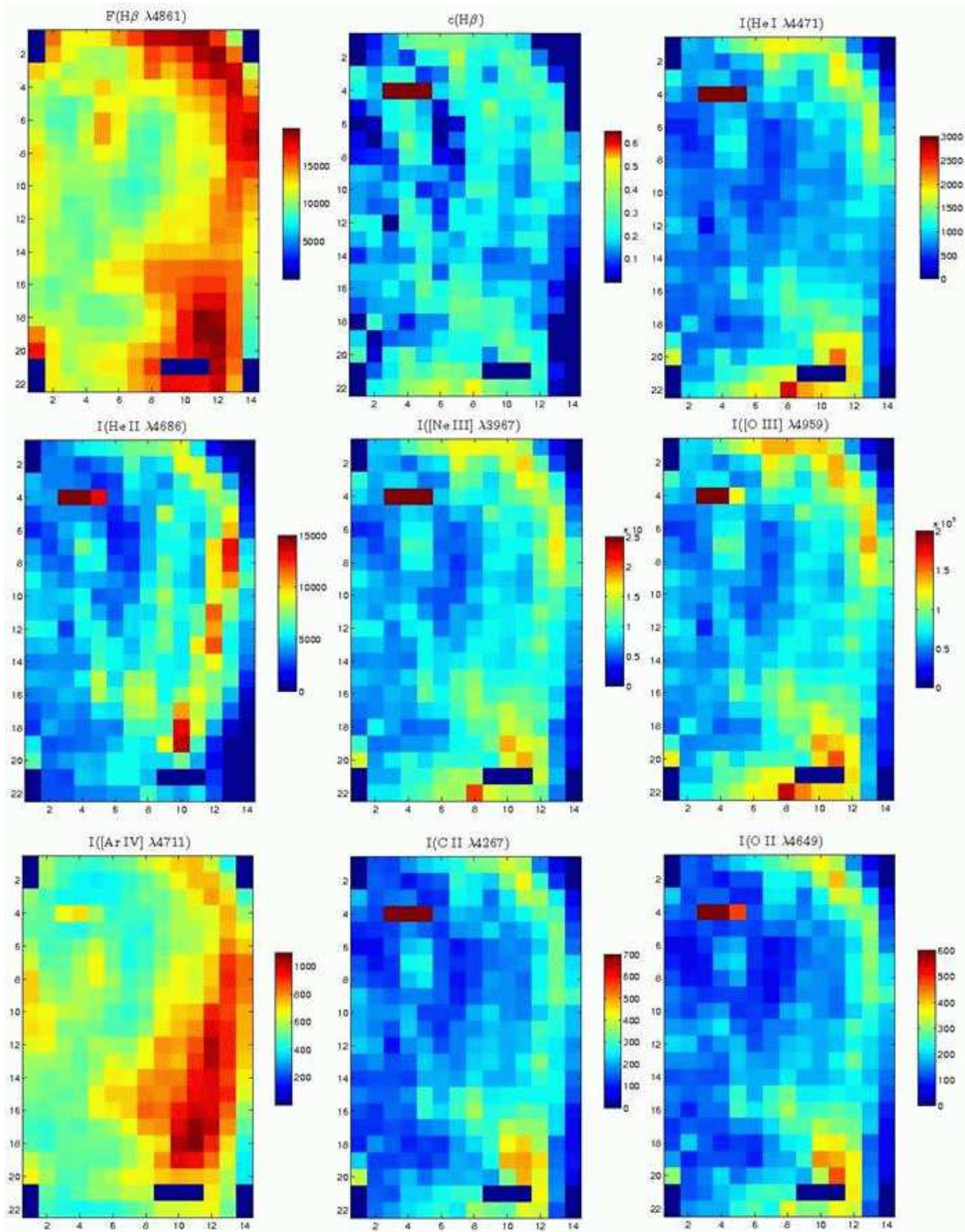, width=14. cm, scale=, clip=, angle=0}

\caption{NGC\,7009 monochromatic line flux ($F$) and dereddened intensity ($I$)
maps (in units of 10$^{-17}$ erg\,cm$^{-2}$\,s$^{-1}$), along with the
extinction constant employed, $c$(\hb). Blank spaxels: three in 2nd row from
bottom (broken Argus fibres); two per corner (fibres reserved for sky
subtraction). Spaxels (3--5, 4) typically suffer from large uncertainties. The
central star position is at coordinates (3, 12). North-east is up (cf.
Fig.\,~3).}
\end{figure*}

\subsection{Emission line maps}

\begin{table*}
\caption{Mean dereddened line ratios, physical conditions and fractional
temperature fluctuation parameters. $F$(\hb) are observed integrated fluxes across the IFUs.
The line intensities are in units of $I$(\hb) $=$ 100. See text for details.}
\begin{tabular}{lccc}
\noalign{\vskip3pt} \noalign{\hrule} \noalign{\vskip3pt}
            & NGC\,5882           & NGC\,6153            & NGC\,7009              \\
\noalign{\vskip3pt} \noalign{\hrule}\noalign{\vskip3pt}

\noalign{\vskip3pt}

$c$(\hb)                    &0.45 $\pm$ 0.24$^a$    & 1.15 $\pm$ 0.12    &0.18 $\pm$ 0.11                    \\
$F$(\hb) (erg\,cm$^{-2}$\,s$^{-1}$)    &2.66 $\cdot$ 10$^{-11}$   &3.12 $\cdot$ 10$^{-12}$     &3.77 $\cdot$ 10$^{-11}$\\
\noalign{\vskip2pt}
$I$(\hei\ $\lambda$4471)    &5.47 $\pm$ 0.23    &6.17 $\pm$ 0.34    &4.90 $\pm$ 0.48               \\
$I$(\heii\ $\lambda$4686)   &4.67 $\pm$ 3.27    &17.4 $\pm$ 8.68    &25.60 $\pm$ 8.27               \\
$I$(\cii\ $\lambda$4267)    &0.491 $\pm$ 0.071  &2.95 $\pm$ 0.58    &1.04 $\pm$ 0.41           \\
$I$(\oii\ $\lambda$4089)    &0.203 $\pm$ 0.108  &0.679 $\pm$ 0.173  &0.343 $\pm$ 0.098         \\
$I$(\oii\ $\lambda$4649)    &0.433 $\pm$ 0.066  &1.88 $\pm$ 0.41    &0.882 $\pm$ 0.152         \\
$I$(\fneiii\ $\lambda$3967) &30.09 $\pm$ 1.77    &32.70 $\pm$ 1.68  &44.15 $\pm$ 6.71           \\
$I$(\foiii\ $\lambda$4959)  &342.3 $\pm$ 11.2    &291.6 $\pm$ 14.2  &372.3 $\pm$ 25.4           \\
$I$(\fariv\ $\lambda$4711)  &1.97 $\pm$ 0.33   &2.66 $\pm$ 0.75     &5.04 $\pm$ 0.88           \\
\fariv\ $\lambda$4711/$\lambda$4740  & 0.867 $\pm$ 0.085     & 0.972 $\pm$ 0.086    & 0.896 $\pm$ 0.034    \\
\foiii\ $\lambda$4959/$\lambda$4363  & 250 $\pm$ 5           & 274 $\pm$ 47       & 187 $\pm$ 18          \\
\noalign{\vskip2pt}
\eld\ (\cmt)         & 5160 $\pm$ 1200     & 3650 $\pm$ 1400      & 4780 $\pm$ 490       \\
\elt\ (K)            & 9520 $\pm$ 190      & 9270 $\pm$ 410       & 10\,340 $\pm$ 380        \\
$T_{0,A}$ (K)        &9560 $\pm$ 2260      &9180 $\pm$ 11290      &10\,270 $\pm$ 1970          \\
$t^2_A$              &0.00034 $\pm$ 0.00021 &0.00180 $\pm$ 0.00234 &0.00075 $\pm$  0.00016     \\

\noalign{\vskip3pt} \noalign{\hrule}\noalign{\vskip3pt}
\end{tabular}
\begin{description}
\item[$^a$]  These are the uncertainties associated with the variation of the
quantity across the IFU region (the rms deviation from the mean). They cannot
be directly compared to the formal error on an integrated flux measurement for
the same region: the uncertainties on the total integrated fluxes and ratios
are $>$ 10 times smaller.

\end{description}
\end{table*}

\subsubsection{NGC~5882}

In Fig.\,~7 a set of NGC\,5882 maps is presented showing the observed \hb, the
logarithmic extinction at \hb, and a sample of representative dereddened
spectral maps in the light of \hei\ $\lambda$4471, \heii\ $\lambda$4686,
\foiii\ $\lambda$4959, \fneiii\ $\lambda$3967, and the recombination lines
\cii\ $\lambda$4267 and \oii\ $\lambda$4649. The top two rows of the array
appear very bright in the `unmasked' \hb\ image: this does not appear in images
of the other two PNe [except perhaps in spaxels (3--5, 22) for the brightest
emission lines of NGC\,6153] and was judged to be a combination of the effects
of differential atmospheric extinction at the edge of the field of view and the
fact that the line flux there shows quite a steep gradient -- this is also
evident from a comparison of our line map with the {\it HST} F555W image of the
PN which shows a confluence of clumped, relatively high surface brightness
emission in that area (Corradi et al. 2000). Such a gradient at the edge of the
array could introduce errors in ratios of lines originating from the different
LR02 and LR03 gratings and so these two rows were masked out and are not
considered in the physical analysis of NGC\,5882 that follows. We also decided
to exclude the faint PN halo from a detailed analysis as the line S/N ratio
there, especially for the \oii\ ORLs, was much lower than the IFU mean. A mask
based on the H$\beta$ signal-to-noise ratio was therefore applied to exclude
the halo region which appears with zero value spaxel intensities (coloured deep
blue) in the $F$(\hb) map of Fig.\,~7. Mean values of the physical conditions
in the halo are given in Sections~4.2.1 and 4.3.1.

The emission lines peak on the main shell of NGC\,5882 and especially on the
north-west corner which coincides with the bright patch visible on the {\it
HST} image (see inset of Fig.\,~1), our standard measure of comparison. The PN
appears elliptical as in the Hubble image, and the southern portion of its halo
is clearly registered in the `unmasked' \hb\ map. The south segment of its main
shell appears faint in all lines and coincides with a low surface brightness
blister-like feature on the {\it HST} image which extends towards the halo. The
central regions are also much fainter relative to the main shell. The \foiii\
and \fneiii\ maps appear very similar, as do the \cii\ and \oii\ ORL maps.
\heii\ $\lambda$4686 is more concentrated on the west part of the PN than on
the east side, in contrast to the \hei\ $\lambda$4471 emission which traces
almost the full nebular ring. The dust extinction is variable across the
nebula; $c$(\hb) has a mean value of 0.449 $\pm$ 0.034 with an rms deviation of
0.241. Assuming a uniform interstellar extinction screen over the face of the
PN, the central-eastern regions appear to be less dusty than the nebular shell.

\subsubsection{NGC~6153}

In Fig.\,~8 a set of NGC\,6153 maps is presented showing the observed \hb, the
logarithmic extinction at \hb, and a sample of representative dereddened
spectral maps of metallic CELs, the helium recombination lines and the \cii\
and \oii\ ORLs. The Argus IFU targeted the brightest portion of the south-east
quadrant of the nebula. The most striking feature in the maps is the high
surface brightness patch peaking at around spaxel (8, 17) and which is clearly
identified with the bright area in the south-eastern quadrant of the {\it HST}
image (cf. Fig.\,~2). This area seems to have lower than average dust
extinction associated with it, at $c$(\hb) $\simeq$ 0.9 -- the extinction is
relatively higher towards the inner part of the nebula, and in the vicinity of
the central star, with values closer to 1.3. The variations may not be very
significant statistically; $c$(\hb) has a mean value of 1.146 $\pm$ 0.180 with
an rms deviation of 0.120. A lower surface brightness discontinuity is present
in the bright area at coordinates $\sim$ (5, 14) in all emission lines except
\fariv\ $\lambda$4711 (which appears bright over a large region). This
discontinuity can be identified with the `thinning' of the bright patch towards
the north-east, something also seen in the inset of Fig.\,~2. \hei\
$\lambda$4471 covers the same area of the IFU as \hb. The \heii\ $\lambda$4686
peak does not coincide with the \hb-bright region but lies interior to it, by 2
spaxels, towards the inner nebula where the plasma ionization state is higher
-- there is also a local \heii\ enhancement near the PN nucleus. The \fneiii\
and \foiii\ maps are very similar and the lines peak at the same spots. The
\cii\ and \oii\ ORL maps are also very similar and show peaks near the \hb\
maximum, but also local peaks in the vicinity of the central star. Spaxels
(3--5, 22) for the brightest emission lines of this PN appear with higher
values than the mean, and may suffer to some unspecified degree from the same
problem as row Y $=$ 22 of NGC\,5882. At the same time, the inset of Fig.\,~2
shows that there is a local flux enhancement within that IFU area.

\subsubsection{NGC~7009}

In Fig.\,~9 we present a similar set of maps for NGC\,7009 as for NGC\,6153.
The Argus array targeted the northern part of the inner shell of the nebula and
the long IFU axis lies in approximately the north-east to south-west direction
(cf. Fig.\,~3). The most conspicuous feature in all maps is the bright arc that
runs from top to bottom and shows high surface brightness spots at either end
-- this is mostly evident in the light of \hb\ but is also present in other
wavelengths too. It can be safely identified with the rim of the inner bipolar
nebular shell visible in the inset of Fig.\,~3. The rim appears fainter in the
middle in all lines, except \heii\ $\lambda$4686 and \fariv\ $\lambda$4711 --
the dust extinction also drops there. The extinction constant displays some
variability with a mean value of 0.183 $\pm$ 0.084 and an rms deviation of
0.111. The \fariv\ line peaks at the lower segment of the rim, whereas \heii\
shows high surface brightness spots in three positions along it.
Characteristically, these spots are all shifted by one spaxel\footnote{The
seeing was less than 1.5 spaxels wide during these observations.} towards the
direction of the PN nucleus, relative to local enhancements of the \hb\
emission, and should correspond to higher ionization zones perhaps associated
with the star-facing side of an expanding shell, that is, zones that lie deeper
in the nebula from the edge of the main ionization front. The \cii\
$\lambda$4267 and \oii\ $\lambda$4649 ORL maps have similar appearances and the
lines also peak at the lower part of this nebular boundary.

\setcounter{figure}{9}
\begin{figure*}
\centering

\epsfig{file=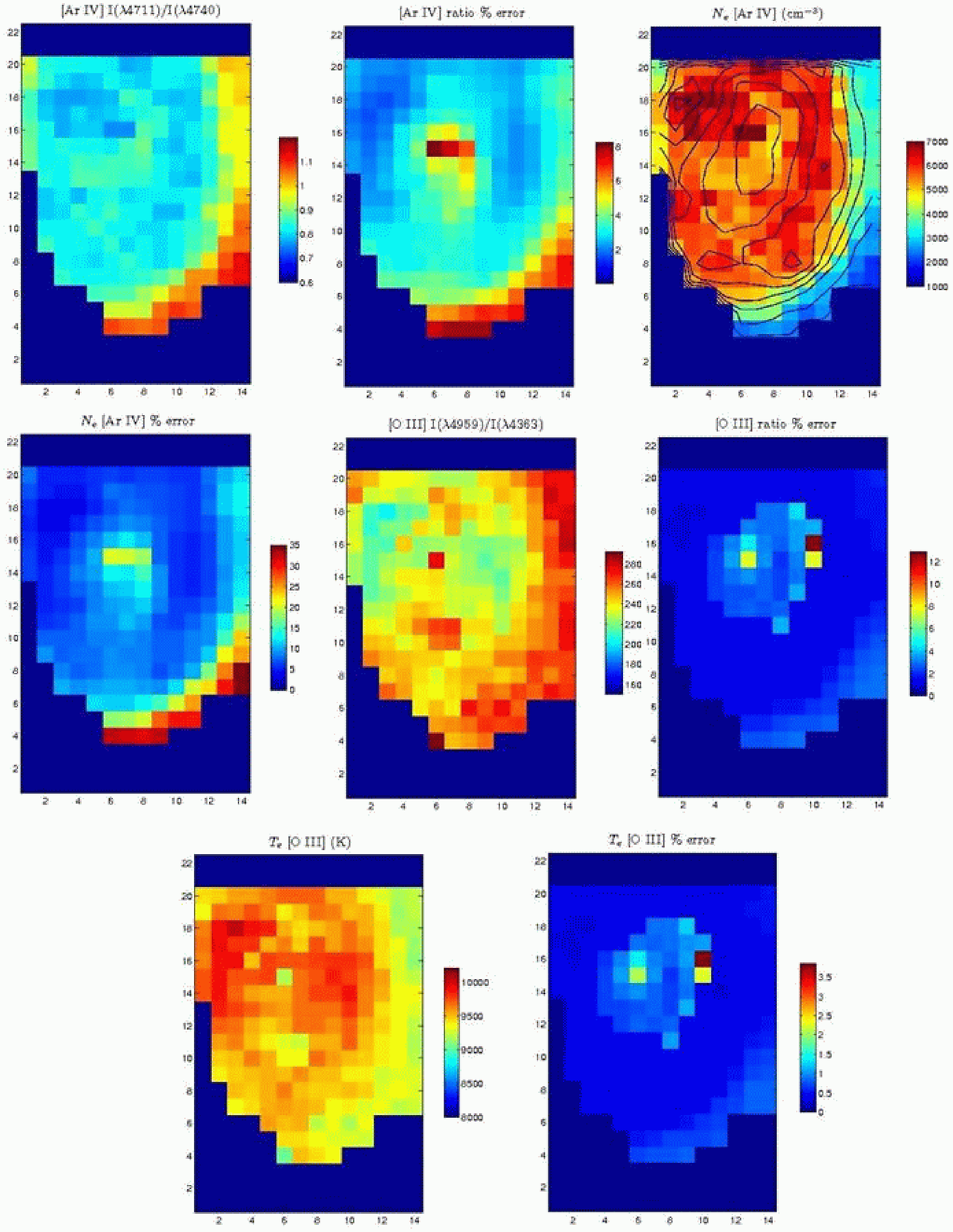, width=14. cm, scale=, clip=, angle=0}

\caption{NGC\,5882: maps of the dereddened \fariv\ and \foiii\ line ratios and
respective errors and of the derived electron density and temperature (and
respective errors). The density map shows overplotted $F$(\hb) isocontours.}
\end{figure*}

\subsection{Plasma physical conditions}

The dereddened emission line maps discussed above were used to compute line
ratio maps and hence the electron temperature (\elt) and density (\eld), and
heavy ion abundances in the PNe relative to hydrogen. The nebular to auroral
\foiii\ $\lambda$4959/$\lambda$4363 ratio was used to derive \elt, and the
\fariv\ $\lambda$4711/$\lambda$4740 ratio was used to derive \eld. Atomic data
for \foiii\ and \fariv\ were taken from Mendoza (1983) and Zeippen et al.
(1987) respectively. Mean values of dereddened ratios of lines used in this
study, and the corresponding plasma conditions for the PNe are reported in
Table~2. These values are in very good agreement with results from the
long-slit surveys.

Table~2 also tabulates the fractional mean-square temperature fluctuation
parameters, $t^2_A$, and average electron temperatures, $T_{0,A}$, across the
IFU area for each PN (that is, in the plane of the sky). We have used equations
11 and 12 of Mesa-Delgado et al. (2007) for the derivation of these quantities
(see also Rubin et al. 2002), based on the observed (\elt, \eld) spatial
distributions. Very small values of the $t^2_A$ parameter in the plane of the
sky are found in all three nebulae. The large uncertainties attached to
($t^2_A$, $T_{0,A}$) for NGC\,6153 are dominated by the relatively large
uncertainties in the per spaxel determination of \eld\ for this nebula. In the
case of NGC\,7009 our $t^2_A$ estimate of 0.00075 is about five times smaller
than that derived by Rubin et al. (2002) from {\it HST}/WFPC2 imagery of the
whole nebula in the \foiii\ lines. These results do not preclude the existence
of large scale temperature variations along a given line of sight. For
instance, Barlow et al. (2006) have presented preliminary evidence for large
line-of-sight \elt(\foiii) variations in NGC\,6153; however, as discussed in
previous works, such large scale temperature gradients cannot be the main cause
of the abundance discrepancy problem (cf. Section~1). Furthermore, Barlow et
al. (2006 and in preparation) discussed new evidence that \oii\ ORLs in
NGC\,6153 and 7009 arise from cool plasma, based on their narrower line widths
compared to that of the \foiii\ $\lambda$4363 line.

\setcounter{figure}{10}
\begin{figure*}
\centering

\epsfig{file=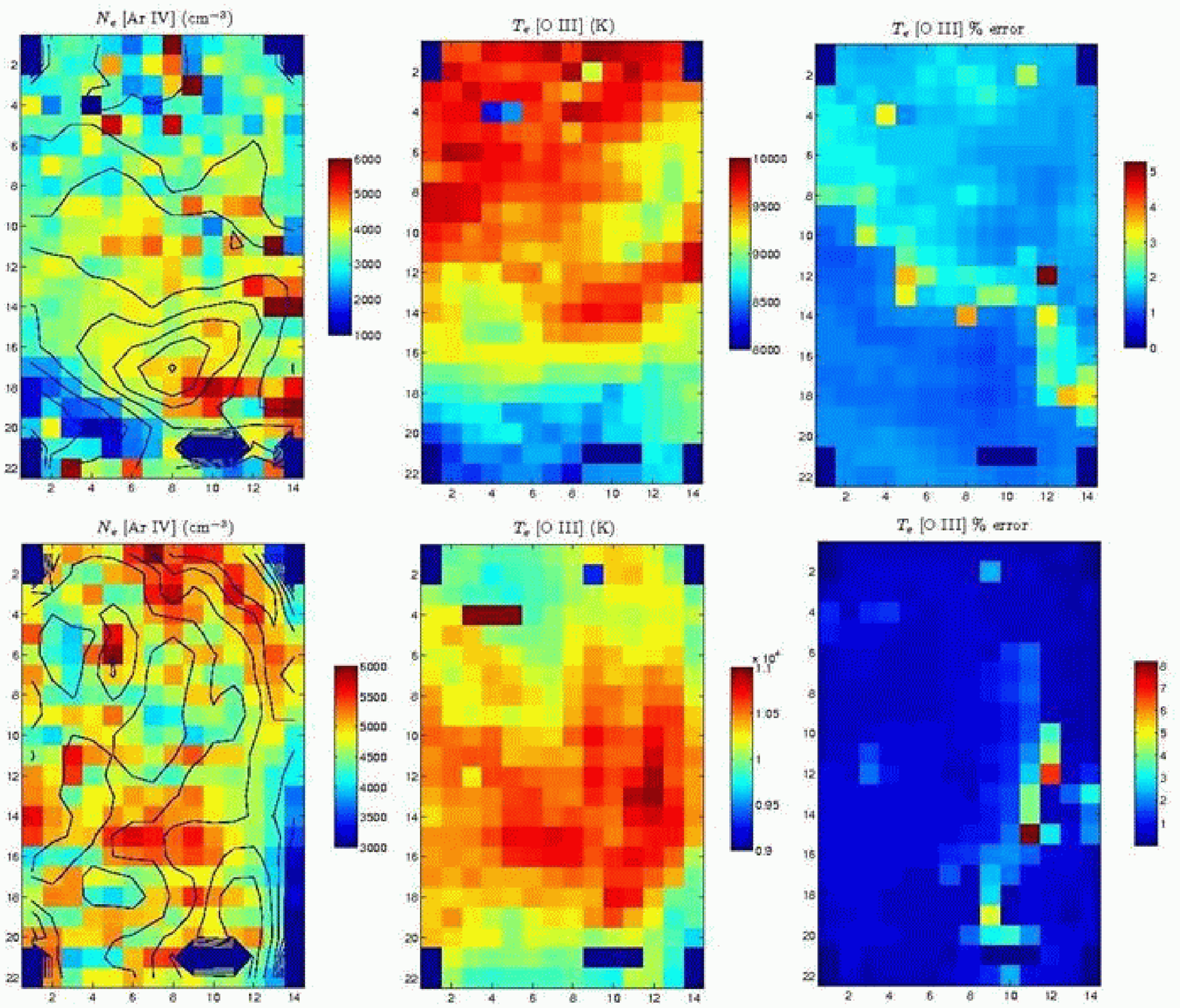, width=14. cm, scale=, clip=, angle=0}

\caption{Electron density and temperature maps for NGC\,6153 (top) and
NGC\,7009 (bottom). The density maps show overplotted $F$(\hb) isocontours.
Spaxels (3--5, 4) are not considered in the analysis.}
\end{figure*}

\subsubsection{NGC~5882}

In Fig.\,~10 maps for NGC\,5882 of the dereddened \fariv\ and \foiii\
diagnostic ratios are presented, along with their respective uncertainties
determined from the line fitting (which incorporate the uncertainties derived
from the pipeline processing), and the corresponding nebular \eld\ and \elt\
maps and their errors. The uncertainties on the \fariv\ ratio are of the order
of 10 per cent, whereas those on the \foiii\ ratio are typically smaller. The
uncertainties on the \eld\ map are of the order of 10 per cent or less along
the bright PN shell, but increase in lower surface brightness regions (towards
the centre, and towards the south-east), where the \fariv\ lines are weaker.
The mean density is 5160 $\pm$ 465 \cmt\ with an rms deviation of 1200 \cmt.
The \elt\ uncertainties are of the order of five per cent or less. The mean
temperature is 9520 $\pm$ 50\,K with an rms deviation of 190\,K. The eastern
part of the PN has the lowest (\elt, \eld). The observed range in temperature
is $\sim$ 1000\,K; a couple of local maxima are seen, the most conspicuous of
which coincides with the \hb-bright patch on the northwest part of the nebular
shell. The mean electron temperature and density in the masked halo region,
contained in the lower part of the IFU aperture [86 spaxels excluding (3--4,
4)], are somewhat lower than in the main shell, at 9380 $\pm$ 150\,K and 2530
$\pm$ 600 \cmt\ respectively.

\subsubsection{NGC~6153}

The \eld\ and \elt\ maps of NGC\,6153 (with the accompanying \elt\
uncertainties only) are presented in Fig.\,~11 (top). This PN shows a
temperature range across the IFU of $\sim$ 1800\,K, with a mean of 9270 $\pm$
150\,K and an rms deviation of 410\,K. The demarcation line running diagonally
across the error map is caused by the transition from single Gaussian profile
fits to the \foiii\ $\lambda\lambda$4363, 4959 lines (lower left) to double
peaked profiles (upper right). The errors are typically larger across the
transition zone because of the increased uncertainty of fitting marginally
resolved lines with double profiles. Where double Gaussian profiles were fitted
the combined flux of both was used in the analysis. The spectral resolution was
high enough to allow a separate \elt\ determination for both nebular velocity
components but, as the line S/N ratio was not sufficiently high, this was not
attempted.

There is a distinct negative temperature gradient from the nucleus towards the
outer nebula (with increasing Y spaxel coordinates out to an offset of 10
arcsec from the nucleus; see also Fig.\,15b). The highest values are observed
near the nucleus, i.e. around spaxel (9, 3), but also in the neighbourhood of
(2, 7). A \elt\ gradient of very similar magnitude along the minor axis of the
PN was also observed by Liu et al. (2000), whose fixed long-slit position is
contained within the IFU orientation (the IFU long-axis covers roughly half the
extent over which those authors performed their spatial analysis). Our
observations further resolve distinct local minima such as those at $\sim$ (13,
8). Liu et al. concluded that the major part of the \elt\ gradient cannot be
due to contamination of the $\lambda$4363 line from recombination excitation.
In Sections~5.4 and 6 the temperature gradient is discussed in the light of the
inhomogeneous dust distribution and the higher dust extinction in the inner
nebula.

From our analysis the \eld\ surface variation in NGC\,6153 does not show a
clear correlation with distance from the nucleus. The range in \fariv\ density
is $\sim$ 5000 \cmt\ with a mean of 3650 $\pm$ 1520 \cmt\ and an rms deviation
of 1400 \cmt. There is some evidence for slightly higher densities over the
\hb\ bright patch, and lower densities at spaxels beyond its outer edge [at
$\sim$ (3, 19); bottom-left IFU region). A similar rise in density was also
seen by Liu et al. roughly correlating with \hb\ surface brightness. No firm
correlation between \hb\ and \eld\ can be however established from the present
analysis as the density uncertainties per spaxel are fairly high.

\subsubsection{NGC~7009}

This nebula shows a mean \elt\ of 10\,340 $\pm$ 90\,K with an rms deviation of
380\,K and a range across the IFU of 1700\,K. In contrast to NGC\,6153, the
temperature in NGC\,7009 (Fig.\,~11, bottom) is not the highest close to the
nucleus at map coordinates (3, 12), but at a position $\sim$ 5 arcsec north
from it at (12, 13). That position coincides with the \heii-bright high
excitation zone discussed in Section 4.1.3 (where the \hepp/He ionization
fraction shows a local maximum; see also Fig.\,~14, bottom row), and marks the
edge of the ionization front of the inner nebular bubble. The \eld\ surface
variation across the IFU shows enhancements in the top and centre-bottom
regions, approximately corresponding to local \hb-bright spots; it is lower
over spaxels (14, 13--20) where the \hb\ surface brightness is low. The mean
electron density across the IFU is 4780 $\pm$ 320 \cmt\ with an rms deviation
of 490 \cmt\ and a range of $\sim$ 3000 \cmt.

\subsection{Chemical abundances}

In this section maps of the chemical abundances for the PNe are discussed. The
\opp/\hp\ abundance ratio for all nebulae was derived from the \foiii\
$\lambda$4959 forbidden line, and also from the \oii\ $\lambda$4089 3d--4f and
$\lambda$4649 3s--3p ORLs. The \cpp/\hp ratio was derived from the \cii\
$\lambda$4267 ORL. The \hep/\hp and \hepp/\hp abundance ratios were derived
from the \hei\ $\lambda$4471 and \heii\ $\lambda$4686 recombination lines
respectively; the sum of the two yielded total He/H abundance ratios for the
three nebulae. The \elt(\foiii) and \eld(\fariv) maps discussed above were
adopted in all cases. The reader is referred to Tsamis et al. (2003b, 2004) for
details on the method of calculating ionic abundance ratios based either on
heavy element CELs or helium and heavier element ORLs. Effective recombination
coefficients are taken from: Storey (1994) for \oii\ $\lambda$4649; Liu et al.
1995 for \oii\ $\lambda$4089; Davey, Storey \& Kisielius (2000) for \cii\
$\lambda$4267 (including both radiative and dielectronic processes); Storey \&
Hummer (1995) for \hei\ and \heii. Table~3 summarizes the mean values from the
maps presented in Figs.\,~12--14, including the abundance discrepancy factors
for \opp. For all targets, the mean {\it ionic} \cpp/\hp\ and \opp/\hp\ ORL
abundance ratios are higher than the corresponding solar abundances of carbon
and oxygen (e.g. Lodders 2003), by factors of $\sim$ 10 (NGC\,6153), $\sim$ 3
(NGC\,7009), and $\sim$ 2 (NGC\,5882). We posit that these overabundance
factors point towards the presence of heavy element-rich (hydrogen-poor)
regions within the nebulae. The total carbon and oxygen recombination-line
abundances, relative to hydrogen, for these regions should be even higher. The
He/H abundances are also higher than solar for all PNe (by up to 40 per cent in
the case of NGC\,6153). In the presented maps the posited hydrogen-deficient
zones are identified as undulations in the ORL-based diagnostics.

\begin{table*}
\caption{Mean chemical abundances and oxygen abundance discrepancy factors (the
ratio of \opp\ ORL over CEL determinations). The listed uncertainties are the
formal errors on the mean. The rms deviations from the mean (associated with
the variation of the quantity across the IFU region) are quoted in the text.}
\begin{tabular}{llccc}
\noalign{\vskip3pt} \noalign{\hrule} \noalign{\vskip3pt}
            && NGC\,5882           & NGC\,6153            & NGC\,7009              \\

\noalign{\vskip3pt} \noalign{\hrule}\noalign{\vskip3pt}

\noalign{\vskip3pt}

\opp/\hp    &\foiii\ $\lambda$4959 &(4.28 $\pm$ 0.02) $\cdot$ 10$^{-4}$    &(4.09 $\pm$ 0.26) $\cdot$ 10$^{-4}$       &(3.54 $\pm$ 0.03) $\cdot$ 10$^{-4}$                           \\
\noalign{\vskip2pt}
\opp/\hp    &\oii\ $\lambda$4089 &(1.83 $\pm$ 0.21) $\cdot$ 10$^{-3}$    &(6.12 $\pm$ 0.63) $\cdot$ 10$^{-3}$       &(3.12 $\pm$ 0.25) $\cdot$ 10$^{-3}$                                                        \\
ADF(\opp)   &$\lambda$4089/$\lambda$4959 &4.33 $\pm$ 0.49  &16.1 $\pm$ 2.0         &9.10 $\pm$ 0.74          \\
\opp/\hp    &\oii\ $\lambda$4649 &(9.10 $\pm$1.60) $\cdot$ 10$^{-4}$    &(4.17 $\pm$ 0.61) $\cdot$ 10$^{-3}$               &(1.78 $\pm$ 0.13) $\cdot$ 10$^{-3}$                           \\
ADF(\opp)   &$\lambda$4649/$\lambda$4959 &2.14 $\pm$ 0.35  &10.9 $\pm$ 2.3 &5.14 $\pm$ 0.35                 \\
\cpp/\hp    &\cii\ $\lambda$4267 &(4.68 $\pm$ 0.37) $\cdot$ 10$^{-4}$    &(2.81 $\pm$ 0.17) $\cdot$ 10$^{-3}$         &(1.01 $\pm$ 0.05) $\cdot$ 10$^{-3}$                           \\
\hep/H      &\hei\ $\lambda$4471    &0.1042 $\pm$ 0.0023    &0.1182 $\pm$ 0.0062        &0.0950 $\pm$ 0.0040                       \\
\hepp/H     &\heii\ $\lambda$4686 &0.0039 $\pm$ 0.0001      &0.0142 $\pm$ 0.0010               &0.0213 $\pm$ 0.0003                           \\
He/H        & (\hep\ $+$ \hepp)/H &0.1081 $\pm$ 0.0023      &0.1324 $\pm$ 0.0063               &0.1163 $\pm$ 0.0040                           \\

\noalign{\vskip3pt} \noalign{\hrule}\noalign{\vskip3pt}
\end{tabular}
\end{table*}

\setcounter{figure}{11}
\begin{figure*}

\epsfig{file=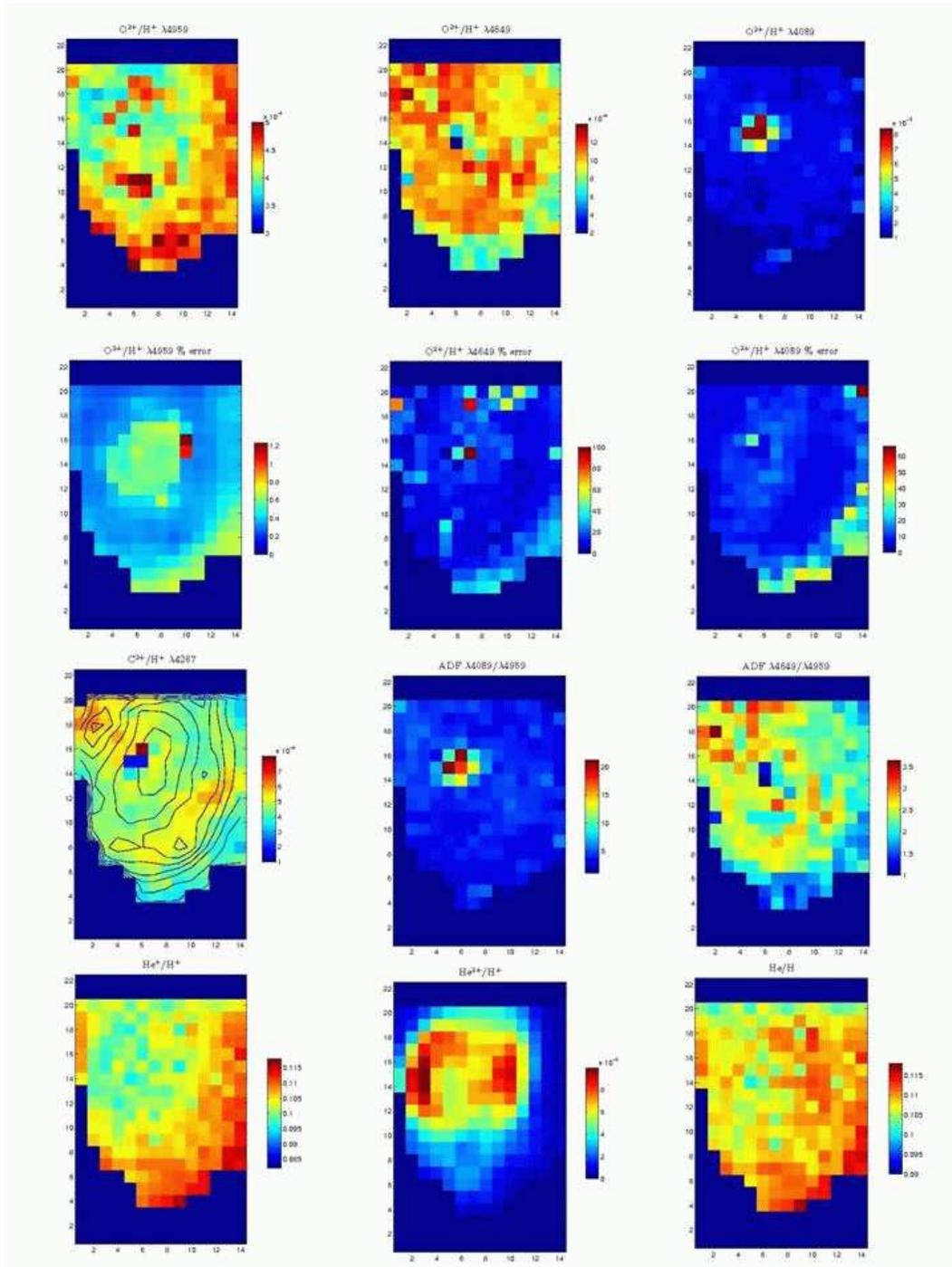, width=14. cm, scale=, clip=, angle=0}

\caption{NGC\,5882: \opp/\hp\ abundance ratios from \foiii\ forbidden and \oii\
recombination lines, \cpp/\hp\ abundances from the \cii\ $\lambda$4267 ORL
[with $F$(\hb) isocontours overplotted], the corresponding oxygen abundance
discrepancy factors, and ionic and total He/H abundances from the helium ORLs.}
\end{figure*}

\setcounter{figure}{12}
\begin{figure*}
\centering \epsfig{file=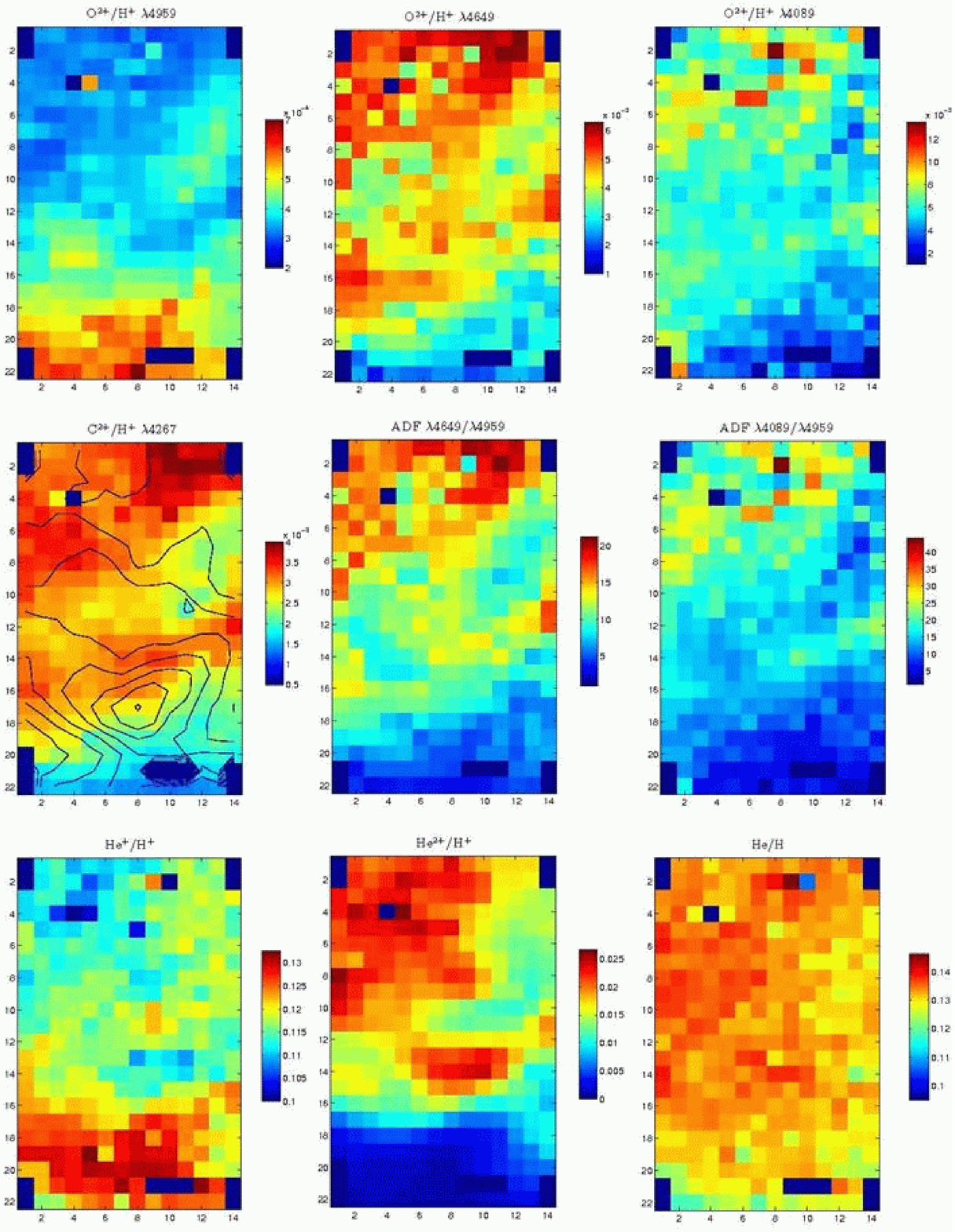, width=14. cm, scale=, clip=, angle=0}

\caption{NGC\,6153: \opp/\hp\ abundance ratios from \foiii\ forbidden and \oii\
recombination lines, \cpp/\hp\ abundances from the \cii\ $\lambda$4267 ORL
[with $F$(\hb) isocontours overplotted], the corresponding oxygen abundance
discrepancy factors, and ionic and total He/H abundances from the helium ORLs.
Spaxels (3--5, 4) are not considered in the analysis.}
\end{figure*}

\setcounter{figure}{13}
\begin{figure*}
\centering

\epsfig{file=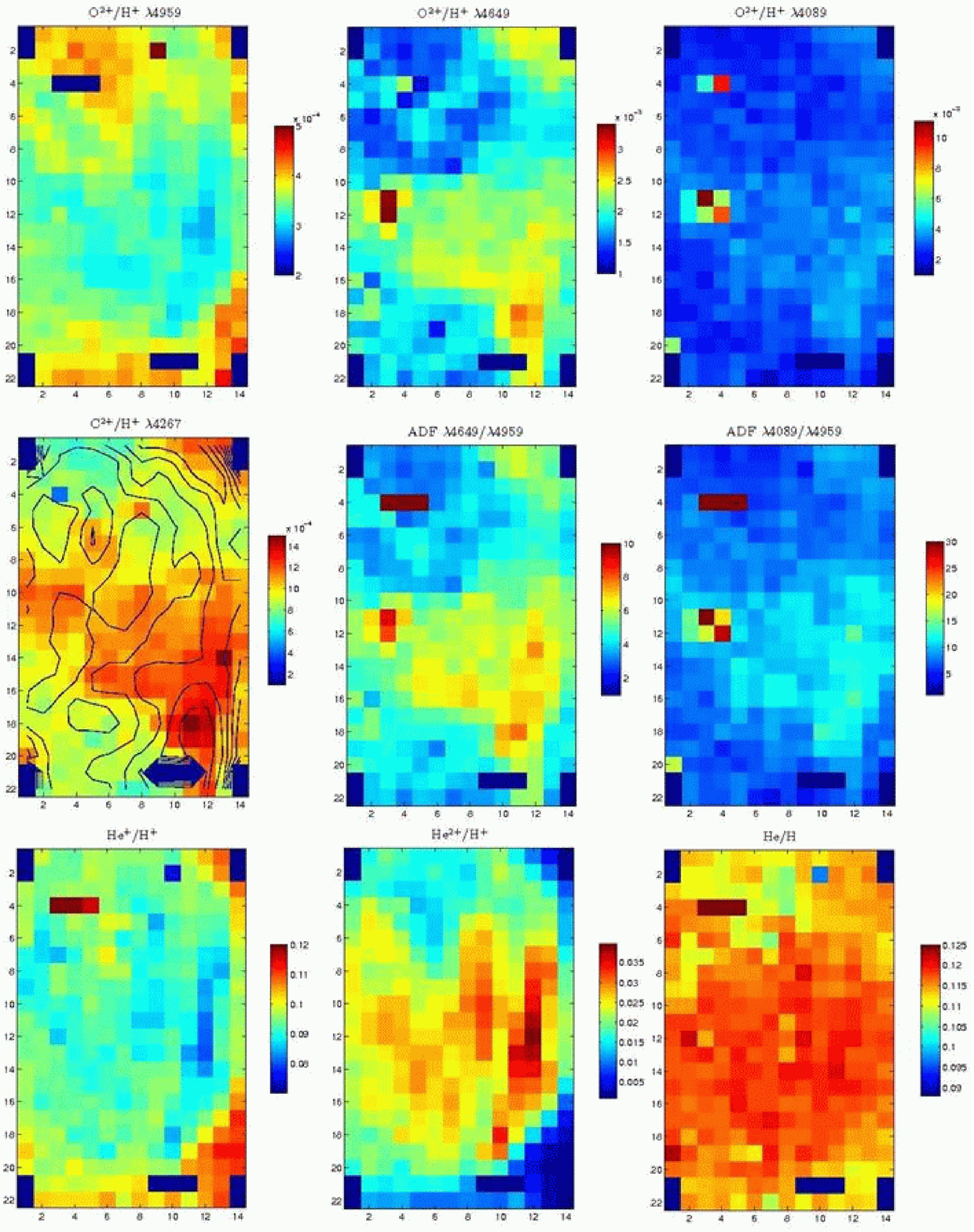, width=14. cm, scale=, clip=, angle=0}

\caption{NGC\,7009: \opp/\hp\ abundance ratios from \foiii\ forbidden and \oii\
recombination lines, \cpp/\hp\ abundances from the \cii\ $\lambda$4267 ORL
[with $F$(\hb) isocontours overplotted], the corresponding oxygen abundance
discrepancy factors, and ionic and total He/H abundances from the helium ORLs.
Spaxels (3--5, 4) are not considered in the analysis.}
\end{figure*}

\subsubsection{NGC~5882}

Abundance maps for NGC\,5882 are presented in Fig.\,~12; the corresponding
error maps are shown for the oxygen diagnostics. The \opp/\hp\ ratio based on
the \foiii\ $\lambda$4959 CEL has a mean value of (4.28 $\pm$ 0.02) $\cdot$
10$^{-4}$ with an rms deviation of 2.97 $\cdot$ 10$^{-5}$. The abundance of
this ion shows a local maximum at the position of the PN nucleus at map
coordinates (6, 15), surrounded by a $\sim$ 19 arcsec$^2$ area where the
abundance is depressed, and two further maxima at positions symmetrically
offset by $\sim$ 2 arcsec from the nucleus on the north-south direction. The
CEL abundance of this ion is further depressed over the bright nebular shell
but increases over the mean value on the eastern side of the PN, where the \hb\
surface brightness and \elt(\foiii) are low. In the masked halo region the
\opp/\hp\ CEL ratio is (4.52 $\pm$ 0.28) $\cdot$ 10$^{-4}$.

The \opp/\hp abundance ratio based on the \oii\ $\lambda$4649 3s--3p ORL shows
a mean of (9.10 $\pm$ 1.60) $\cdot$ 10$^{-4}$ with an rms variation of 1.22
$\cdot$ 10$^{-4}$. The lowest values of the ratio are found in the south-east
corner of the IFU. The ionic abundance based on this line is perhaps slightly
depressed over a four spaxel area in the vicinity of the nucleus with a value
of (6.00 $\pm$ 3.13) $\cdot$ 10$^{-4}$. On the other hand, the \opp/\hp ratio
based on the \oii\ $\lambda$4089 3d--4f line shows a very different surface
distribution relative to the $\lambda$4649 transition. Whereas the mean value
is (1.83 $\pm$ 0.21) $\cdot$ 10$^{-3}$ with an rms deviation of 9.76 $\cdot$
10$^{-4}$, the region in the vicinity of the PN nucleus over a nine spaxel area
shows a three times higher abundance with a value of (5.52 $\pm$ 0.61) $\cdot$
10$^{-3}$. From the long slit surveys it was established that, on average, the
$\lambda$4089 transition typically returns higher \opp/\hp\ ratios than the
$\lambda$4649 transition when both line emissivities are calculated under the
assumption of \elt(\foiii) for their emitting regions. This effect, when
attributed to the different \elt\ dependence of the emissivities of the two
classes of \oii\ transitions (3s--3p {\sl versus} 3d--4f), has been used to
provide evidence for the existence of cold plasma regions in several PNe
consistent with the existence of hydrogen-deficient regions (Tsamis et al.
2004; Liu et al. 2004). In Tsamis et al. (in preparation) we will revisit this
issue and present a detailed analysis of the \oii\ $\lambda$4089/$\lambda$4649
ratios from the current data-set based on the latest effective recombination
coefficients. In the masked halo region the \opp/\hp\ ORL ratio from the
$\lambda$4649 and $\lambda$4089 lines is (6.96 $\pm$ 1.21) $\cdot$ 10$^{-4}$
and (1.37 $\pm$ 0.49) $\cdot$ 10$^{-3}$ respectively, rather consistent with
the values across the main shell of the nebula.

We also present the respective \opp\ ADF maps, that is the
\opp\,$\lambda$4649/\opp\,$\lambda$4959 and
\opp\,$\lambda$4089/\opp\,$\lambda$4959 quantities. These appear different
(caused by the different behaviour of the \opp\ ORL maps discussed above) with
the former showing local maxima along the nebular shell and the latter showing
a sharp maximum near the PN nucleus.

The \cpp/\hp\ ratio derived from the \cii\ $\lambda$4267 ORL shows a mean of
(4.68 $\pm$ 0.37) $\cdot$ 10$^{-4}$ with an rms deviation of 6.82 $\cdot$
10$^{-5}$. Higher values are observed at the northwest and southeast segments
of the \hb-bright shell. The value of the ratio over four spaxels in the
vicinity of the central star is (4.26 $\pm$ 1.11) $\cdot$ 10$^{-4}$ and is
therefore consistent with the mean across the IFU (the strong stellar continuum
in the neighbourhood of the PN nucleus can sometimes introduce higher errors in
the line fitting -- higher spectral resolution data would help in such cases).
In the masked halo region this abundance ratio is (3.88 $\cdot$ 0.42) $\cdot$
10$^{-4}$.

Finally, Fig.\,~12 (bottom) shows the helium abundance maps of NGC\,5882. The
mean He/H ratio is 0.1081 $\pm$ 0.0023 with an rms deviation of 0.0029. The
\hep/\hp\ abundance ratio is depressed in the central and western nebular
regions, but is enhanced in the south and eastern area which is bordering on
the PN halo. In contrast, the \hepp/\hp\ abundance ratio shows two prominent
maxima offset symmetrically on either side of the PN nucleus along the
east-west axis. The \hep/\hp\ ratio is inversely correlated with the \hepp/\hp
ratio across the IFU region (see Table~A1). In the masked halo region the
\hep/\hp\ and \hepp/\hp\ ratios are 0.1142 $\pm$ 0.0056 and (9.14 $\pm$ 6.10)
$\cdot$ 10$^{-4}$ respectively.

\subsubsection{NGC~6153}

Abundance maps for NGC\,6153 are presented in Fig.\,~13. The \opp/\hp\ ratio
based on the \foiii\ $\lambda$4959 CEL has a mean value of (4.09 $\pm$ 0.26)
$\cdot$ 10$^{-4}$ with an rms deviation of 8.51 $\cdot$ 10$^{-5}$. The
forbidden line abundance of this ion shows an inverse correlation with
\elt(\foiii), being high where the plasma temperature is low
(characteristically in the outer nebula mainly; see Section~5) and vice-versa.
In contrast, the \opp/\hp ratio based on the \oii\ $\lambda\lambda$4089, 4649
ORLs generally increases towards the inner nebula, being proportional to the
$I$(\oii)/$I$(\hb) brightness ratio surface distribution. Both of the \oii\
recombination line abundance diagnostics show local maxima in the vicinity of
the PN nucleus, but as in the case of NGC\,5882 (and for the same reasons),
their respective maps have a different appearance. For this reason, the
$\lambda$4089/$\lambda$4959 ADF map shows values as high as $\sim$ 40 and the
$\lambda$4649/$\lambda$4959 ADF map values as high as $\sim$ 20 at spaxels near
the PN nucleus.

In the same vein, the \cpp/\hp\ ratio derived from the \cii\ $\lambda$4267 ORL,
with a mean of (2.81 $\pm$ 0.17) $\cdot$ 10$^{-3}$ and an rms deviation of 5.72
$\cdot$ 10$^{-4}$, strongly peaks near the PN nucleus as well.
Characteristically, both the ionic carbon abundance and the ionic oxygen
abundance (from the \oii\ $\lambda$4649 line) peak in the neighbourhood of
spaxel (10, 2) about 2~arcsec from the nucleus; the \opp/\hp ratio from
$\lambda$4649 has a value of (6.14 $\pm$ 0.81) $\cdot$ 10$^{-3}$ over a three
spaxel area when the mean over the IFU is (4.17 $\pm$ 0.61) $\cdot$ 10$^{-3}$;
this represents a local enhancement of 47 per cent over the mean. Across the
same region the \cpp/\hp\ ORL abundance is (3.92 $\pm$ 0.27) $\cdot$ 10$^{-3}$,
that is 40 per cent higher than the mean. It therefore seems that in the case
of NGC\,6153 the PN nucleus, either in itself or via its hard radiation field,
is strongly implicated in the causes of the abundance discrepancy, and in the
high abundances derived from oxygen and carbon ORLs. The peaking of the \cii\
and \oii\ abundance diagnostics in the central PN regions had also been
inferred from the long-slit analysis (Liu et al. 2000).

In Fig.\,~13 (bottom) we show the helium abundance maps of NGC\,6153. The mean
He/H ratio is 0.1324 $\pm$ 0.0063 with an rms deviation of 0.0039. The
\hep/\hp\ abundance ratio is lower in the inner nebular regions (mid-top IFU
area), but is enhanced over its mean value in the outer nebula. In contrast,
the \hepp/\hp\ ratio is higher in the inner PN (see also Section~5). As with
NGC\,5882, there is an inverse linear correlation between the \hep/\hp and
\hepp/\hp ratios across all spaxels (see Table~A1). The total He/H abundance
possibly shows a small positive gradient towards the inner nebula (top-left IFU
area).

\subsubsection{NGC~7009}

In Fig.\,~14 abundance maps for NGC\,7009 are presented. The \opp/\hp\ ratio
based on the \foiii\ $\lambda$4959 CEL has a mean value of (3.54 $\pm$ 0.03)
$\cdot$ 10$^{-4}$ with an rms deviation of 4.42 $\cdot$ 10$^{-5}$. In this PN
also, the forbidden-line abundance of the \opp\ ion shows an inverse
correlation with \elt(\foiii), being high where the plasma temperature is low
(see Section~5) and vice-versa. The abundance is generally depressed over a
triangular region in the middle of the IFU. In contrast, over the same area the
\oii\ and \cii\ recombination line abundance diagnostics are enhanced. There
are prominent maxima in the \opp/\hp\ ORL ratio in the vicinity of the PN
nucleus near map coordinates (3, 12) where the ratio based on the $\lambda$4649
line has a value of (3.36 $\pm$ 0.62) $\cdot$ 10$^{-3}$ over a two spaxel area;
this constitutes a local enhancement of 89 per cent over the corresponding mean
across the map of (1.78 $\pm$ 0.13) $\cdot$ 10$^{-3}$. The respective ADF maps
show peaks near the nucleus as well. The \cpp/\hp\ ratio peaks close to the
nucleus with a value of (1.14 $\pm$ 0.16) $\cdot$ 10$^{-3}$ over a four spaxel
area, but also near the edge of the \hb-bright rim, some 5 arcsec away from the
central star; it has a value of (1.49 $\pm$ 0.04) $\cdot$ 10$^{-3}$ at
coordinates (11, 18). Its mean value across the IFU is (1.01 $\pm$ 0.05)
$\cdot$ 10$^{-3}$ with an rms deviation of 1.57 $\cdot$ 10$^{-4}$.

In Fig.\,~14 (bottom) the helium abundance maps of NGC\,7009 are shown. The
mean He/H ratio is 0.1324 $\pm$ 0.0063 with an rms deviation of 0.0039. The
\hep/\hp\ abundance ratio is generally low over the triangular area where the
\opp\ and \cpp\ ORL abundances are enhanced, whereas the \hepp/\hp\ ratio is
correspondingly high there, peaking near the rim of the inner nebular bubble.
There is a strong linear inverse correlation between the \hep/\hp and \hepp/\hp
ratios in this nebula too (see Table~A1). The total He/H abundance shows
slightly lower values in the top of the IFU map, but does not vary across the
edge of the inner nebular bubble where the \hepp/He ionization fraction peaks.

\section{Correlations}

In this section we investigate several correlations between physical properties
of the PNe that can be established following this analysis. Evidence for a few
of these existed from the long-slit surveys, whereas others are new. A unique
feature of this study, however, is that the large number of data points ($\sim$
300) per PN allowed us to put all these on a firm footing for the first time.
Moreover, the 2D aspect of this analysis is particularly revealing, and in the
following paragraphs it shall be brought to the fore. We present results for
each nebula individually so as to first highlight the differences and then
address the similarities; NGC\,6153 is examined first.

\subsection{NGC~6153}

\setcounter{figure}{14}
\begin{figure*}
\centering \epsfig{file=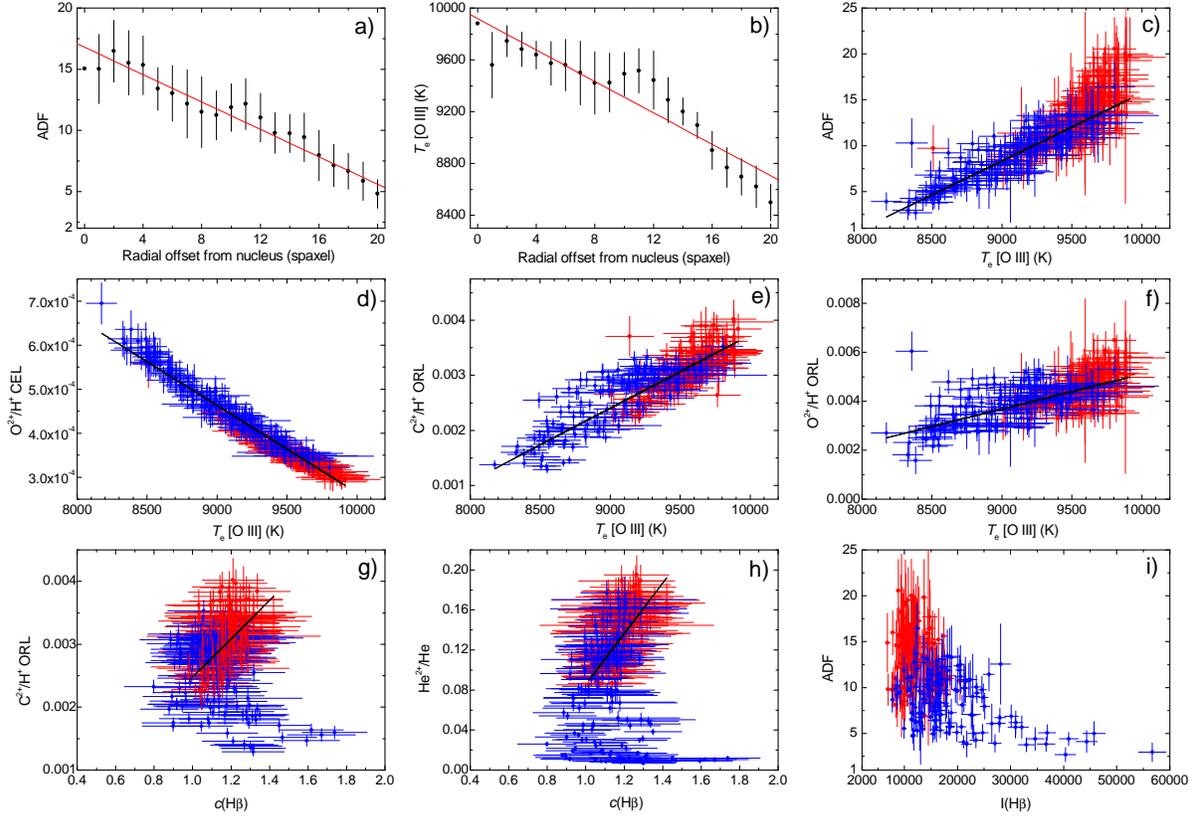, width=13.5 cm, scale=, clip=, angle=-90}
\caption{NGC\,6153: Correlations between various physical quantities: (a) The
\opp\ ADF and (b) the \foiii\ \elt\ radial profiles resulting from binning the
respective Argus maps at equal radii from the central star. The full radial
offset from the PN nucleus (20 spaxels) is 10.4 arcsec; (c)--(i) Red circles
correspond to Argus IFU spaxel coordinates (X, Y) $=$ (1--14, 1--11) and blue
diamonds to coordinates (1--14, 12--22) -- see Fig.\,~8. The solid lines are
weighted fits; (g)--(h) The solid lines are weighted fits to the red data
points only. See text for details.}
\end{figure*}

\setcounter{figure}{15}
\begin{figure*}
\centering \epsfig{file=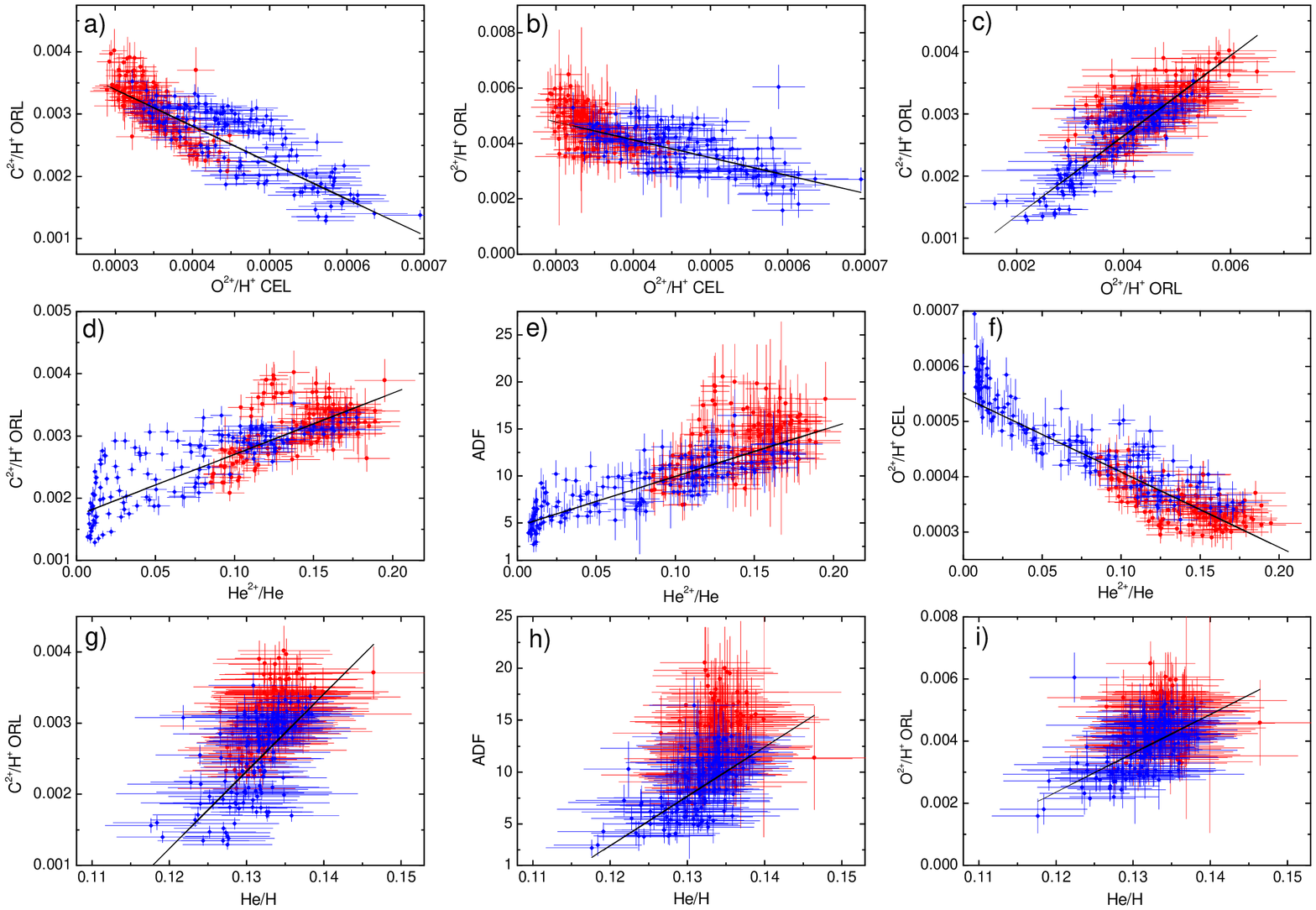, width=13.5 cm, scale=, clip=, angle=-90}
\caption{NGC\,6153: Correlations between various physical quantities with
symbols as in Fig.\,~15; the solid lines are weighted fits. See text for
details.}
\end{figure*}

The \opp\ ADF and the \foiii\ electron temperature for this PN show a very
distinct correlation with distance from the PN nucleus. We have taken the Argus
maps of these physical quantities (Figures 11 and 13) and binned all non-zero
value spaxels at equal radii outwards from the position of the central star
(with a radial increment of one spaxel); in Fig.\,~15a--b the result is shown.
As the \foiii\ temperature rises from the outer to the inner regions of the PN
from $\sim$ 8400 to 10\,000\,K, characteristically the oxygen ADF rises as
well. The quantity plotted in Fig.\,~15a is the $\lambda$4649/$\lambda$4959 ADF
(and is also used in Figs.\,~16--18), but a linear correlation is obtained for
the $\lambda$4089/$\lambda$4959 ADF as well.

Fig.\,~15c shows the relationship between the oxygen ADF and \elt(\foiii) for
the 296 spaxels with a measurement of these quantities. The solid line is an
error-weighted least-squares fit to the data with a correlation coefficient
($r$) of 0.92. In Fig.\,~15d the inverse relationship between the \opp/\hp\
abundance ratio based on the \foiii\ $\lambda$4959 line and \elt(\foiii) is
shown; the correlation is very tight in this case too with a linear correlation
coefficient of $-$0.98. In Figs.\,~15 and 16, black solid lines are
least-squares fits weighted by the errors, red circles denote data from spaxels
with coordinates (from the corresponding Argus maps) of (X, Y) $=$ (1--14,
1--11), whereas blue diamonds correspond to coordinates (1--14, 12--22). This
colour scheme allows us to examine the behaviour of nebular properties in
nebular regions that are near and far out from the central star (red and blue
symbols respectively; cf. also the orientation of the Argus IFU long axis viz.
the position of the PN nucleus in Fig.\,~2). From the juxtaposition of
Fig.\,~15a--b and 15c--d a consistent picture emerges. Regions closer to the
nucleus of NGC\,6153 have higher temperature, lower forbidden line \opp/\hp\
abundance and exhibit higher \opp\ ADFs than regions farther out in the nebula.
Some unavoidable intermixing of the colour-coded data is seen due to projection
(line of sight) effects and/or due to the fact that the steady progression from
lower to higher temperatures has only been highlighted with a two-colour
scheme.

In Fig.\,~15e--f the correlations between the ORL abundance ratios \cpp/\hp\
and \opp\/\hp\ {\sl versus} \elt(\foiii) are plotted; the correlation
coefficients of the error-weighted fits are in this case 0.88 and 0.70
respectively. These plots show that the recombination line abundances of doubly
ionized carbon and oxygen, which are temperature-insensitive to a large degree,
(i) rise towards the central nebular regions, and (ii) are typically the
highest where the forbidden line \foiii\ temperature is highest. The \opp/\hp\
ORL ratio used in the correlations discussed and plotted in this section is
based on the \oii\ $\lambda$4649 line.

We also find a correlation of the \cpp/\hp\ ORL abundance ratio and the
\hepp/He ionization fraction {\sl versus} the extinction constant, $c$(\hb); we
performed least-squares fits for the inner nebula ($\sim$ 148 data points) and
show the results in Fig.\,~15g--h. Fig.\,15i shows that an inverse trend is
established between the oxygen ADF and the dereddened \hb\ flux from each Argus
spaxel. The trend is clearly non-linear. Regions of low surface brightness (in
\hb) that lie in the inner nebula exhibit the highest oxygen ADFs, whereas the
low ADF `tail' corresponds to brighter zones farther out in the nebula (most
points in this tail correspond to spaxels of the last few IFU rows -- large Y
coordinates -- on the far side from the PN nucleus).

The correlations between the ORL abundance ratios \cpp/\hp\ and \opp/\hp\ {\sl
versus} the forbidden-line \opp/\hp\ abundance ratio are shown in
Fig.\,~16a--b, where an inverse trend is seen in both cases with $r$'s of
$-$0.86 and $-$0.69 respectively. These are mainly driven by the very strong
linear correlation between the temperature-dependent forbidden-line \opp/\hp\
ratios and \elt(\foiii) that was discussed earlier. In stark contrast, a
positive correlation between the temperature-insensitive ORL \cpp/\hp\ and ORL
\opp/\hp\ abundance ratios is established with $r$ $=$ 0.87 (Fig.\,~16c). The
fact that the trends between the ORL--CEL and ORL--ORL abundance diagnostics
lie in completely opposite directions shows that (i) {\it vastly different
assumptions govern the physical origins and the metallicity of the gas traced
by ORL diagnostics on the one hand and CEL diagnostics on the other}, and that
(ii) {\it consistent nebular metallicity estimates can in principle only be
derived using pure ORL/ORL or CEL/CEL emission line ratios}.\footnote{This may
not work for helium however; see Section~6.} This further tells us that the
\cpp/\opp\ (pure ORL) abundance ratio is approximately constant throughout the
portion of NGC\,6153 observed in this programme with a mean value of 0.68 $\pm$
0.11.

Fig.\,~16d--e shows that the \cpp/\hp\ abundance ratio and the oxygen ADF are
positively correlated with the \hepp/He ionization fraction with $r$'s of 0.85
and 0.86 respectively. The \hepp/He fraction rises close to the PN nucleus
where the ionization degree of the nebular plasma is higher due to the harder
radiation field, so these plots are again consistent with an origin for the
ORL-emitting heavy element-rich plasma in the inner regions of the nebula. On
the other hand an inverse linear trend is established between the \opp/\hp CEL
abundance ratio and \hepp/He ($r$ $=$ $-$0.87; Fig.\,~16f).

In Fig.\,~16g--h the correlations between the \cpp/\hp\ ORL abundance, the
oxygen ADF, and the \opp/\hp\ ORL abundance {\sl versus} the He/H abundance
ratio are shown. Positive, but weaker, trends are seen in all cases with $r$'s
of 0.68, 0.59, and 0.55 respectively. These plots also reveal that typically
regions in the inner nebula show a somewhat higher helium abundance (but note
that this large He/H is challenged in Section~6). Finally, we do not find any
obvious trends of the \cpp/\hp\ or \opp/\hp\ ORL abundances (or of the \opp\
ADF) with electron density. The detailed parameters of our regression analysis
for this nebula are listed in Table~A1. Parameters of fits obtained when the
\opp/\hp\ ORL abundance is based on the $\lambda$4089 line are also listed in
Table A1 (the corresponding plots are not shown): the fits are all in the same
direction as previously, and of similar strength, but with a slightly larger
scatter [as the per spaxel uncertainties on $I$($\lambda$4089) are somewhat
higher].

\subsection{NGC~7009}

\setcounter{figure}{16}
\begin{figure*}
\centering \epsfig{file=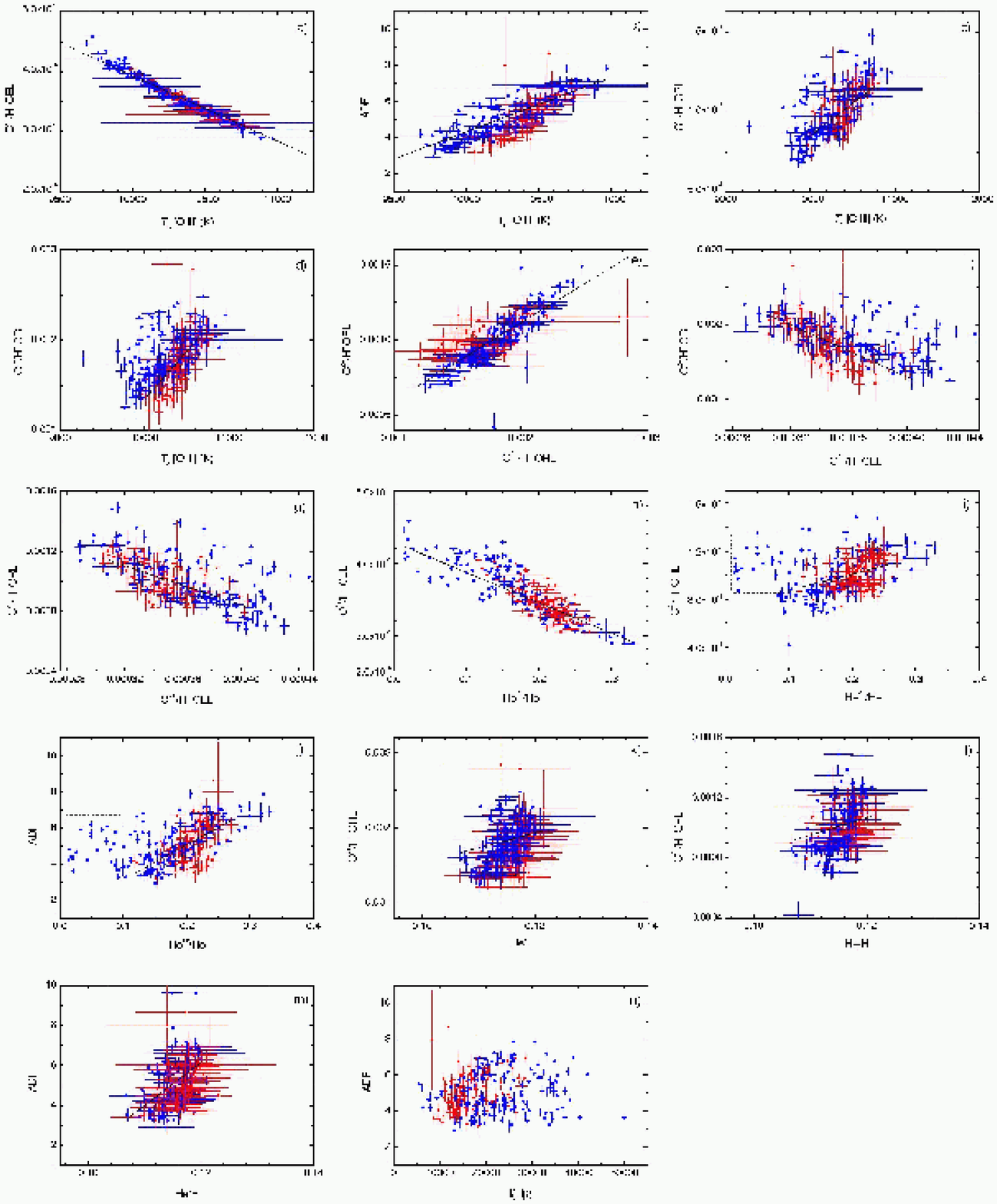, width=15 cm, scale=, clip=, angle=0}
\caption{NGC\,7009: Correlations between various physical quantities. The solid
lines are weighted fits. Red circles correspond to Argus IFU spaxel coordinates
(X, Y) $=$ (1--9, 5--19) and blue diamonds correspond to spaxels outside this
area (see Fig.\,~9); (f, g) and (i, j): the solid lines are fits to the red
data points only. See text for details.}
\end{figure*}

\setcounter{figure}{17}
\begin{figure*}
\centering \epsfig{file=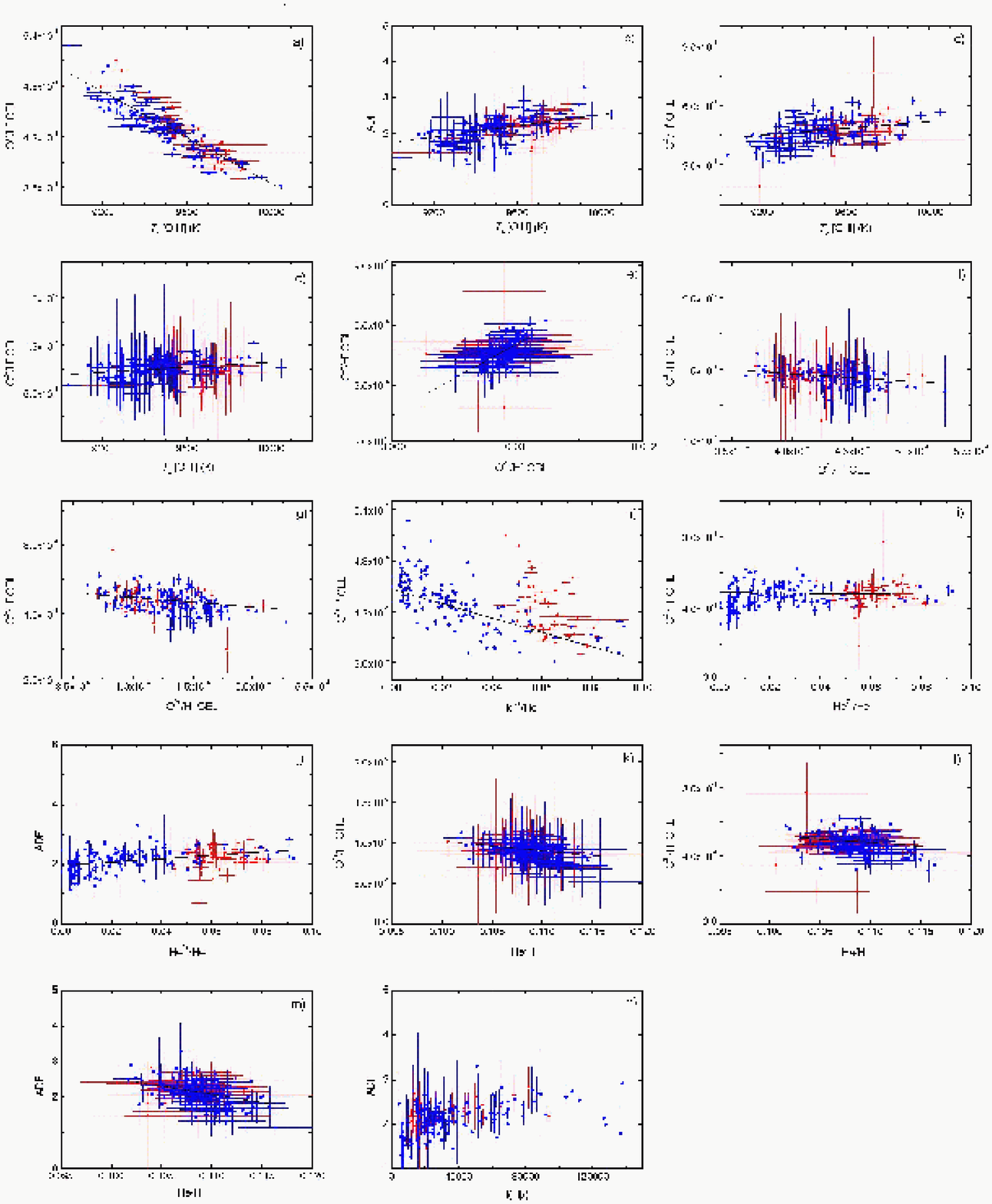, width=15 cm, scale=, clip=, angle=0}
\caption{NGC\,5882: Correlations between various physical quantities. Red
circles correspond to Argus IFU spaxel coordinates (X, Y) $=$ (4--10, 11--18)
and blue diamonds correspond to spaxels outside this area (see Fig.\,~7). The
solid lines are linear fits weighted by the errors (only through the blue data
points in panel `h'). See text for details.}
\end{figure*}

In this section we examine whether the various correlations between diagnostics
that were established for NGC\,6153 hold for this nebula too. In order to check
for any markedly different behaviour between nebular regions the data from the
Argus maps have been separated according to the following scheme: in the
following plots, data from spaxels with coordinates -- see Fig.\,~9 -- (X, Y)
$=$ (1--9, 5--19) are represented by red circles, whereas data points outside
this rectangular area are marked by blue diamonds. The red symbols thus
correspond to a nebular area of lower mean surface brightness (in dereddened
\hb) and of somewhat closer proximity to the PN nucleus than the blue symbols
(cf. Fig.\,~3). The blue symbols comprise the zones of higher surface
brightness and specifically the bright \hb\ rim of the inner PN shell which
runs from top to bottom in the \hb\ image of the nebula (cf. Fig.\,~9 and the
inset of Fig.\,~3 where this sharp boundary is clearly seen in the {\it HST}
image). This nebula is however larger than NGC\,6153 and the Argus IFU targeted
the central regions of the inner bipolar shell; hence the adopted grouping of
spaxels in the plots that follow may not be as clear-cut regarding their
distance from the nucleus as it was for NGC\,6153 where the IFU was both more
favourably placed and captured a larger portion of the PN.

In Fig.\,~17a the inverse relationship between the \opp/\hp\ abundance ratio
based on the \foiii\ $\lambda$4959 line and \elt(\foiii) is plotted for the 296
spaxels with a measurement of these quantities. The solid line is an
error-weighted least-squares fit to the data with a correlation coefficient of
$-$0.94. The positive trend between the oxygen ADF and \elt(\foiii) is shown in
Fig.\,~17b and has a correlation coefficient of 0.78.

In Fig.\,~17c--d the positive correlations between the ORL abundance ratios
\cpp/\hp\ and \opp/\hp\ {\sl versus} \elt(\foiii) are shown; the $r$'s of the
error-weighted fits through the 135 red data points of the inner nebular field
are in this case 0.66 and 0.70 respectively; they are 0.50 and 0.35,
respectively, when fitting all the data. These plots show that the
recombination line abundances of doubly ionized carbon and oxygen for NGC\,7009
follow the same trend with forbidden line temperature as was the case of
NGC\,6153. The \opp/\hp\ ORL ratio used in the correlations plotted in this
section is based on the \oii\ $\lambda$4649 line.

In Fig.\,~17f--g we plot the correlations between the ORL abundance ratios
\opp/\hp\ and \cpp\/\hp\ {\sl versus} the forbidden-line \opp\/\hp\ abundance
ratio, where an inverse trend is seen in both cases with $r$'s of $-$0.68 and
$-$0.63 respectively for fits to the red data points. When fits to all data are
performed the coefficients are less than 0.50 due to the larger scatter of some
of the blue data points which correspond to nebular zones of higher mean
surface brightness. The general behaviour of the trends is the same as for
NGC\,6153. A positive correlation between the temperature-insensitive ORL
\cpp/\hp\ and ORL \opp/\hp\ abundance ratios is established ($r$ $=$ 0.86;
Fig.\,~17e). The \cpp/\opp\ (pure ORL) abundance ratio is approximately
constant throughout the field but there are a few (red) points a factor of
$\sim$1.27 higher (but within 2$\sigma$) than the mean ratio of 0.57 $\pm$
0.05.

Fig.\,~17i--j shows that the \cpp/\hp\ ORL abundance ratio and the \opp\ ADF
are positively correlated with the \hepp/He ionization fraction with $r$'s of
0.35 and 0.55 respectively (for fits through the red data points only). The
\opp/\hp CEL ratios are inversely correlated with the helium ionization state
(the weighted fit to all data has an $r$ of $-$0.84). There is more scatter in
these plots than was the case for NGC\,6153: the cluster of low ionization
points within the dashed squares in Fig.\,~17i--j (at ADF $\sim$ 5) come from
the lower-right corner of the corresponding \hepp/He map and also from IFU row
Y $=$ 14; they are also of relatively high \opp/\hp (forbidden-line) abundance.
The main trends identified in these plots are in the same vein as for
NGC\,6153, with high ionization gas exhibiting higher oxygen ADFs and higher
\cpp/\hp\ ORL abundance ratios than lower ionization, high \opp/\hp CEL
regions.

Fig.\,~17k--m shows the ORL \opp/\hp\ and \cpp\/\hp\ abundance ratios and the
\opp\ ADF {\sl versus} the He/H abundance ratio. Weak positive trends are seen
in all cases -- for the \opp\ ADF vs. He/H linear fit, $r$ $=$ 0.58. The trends
are in the same direction as in the case of NGC\,6153. We further plot the
\opp\ ADF {\sl versus} the dereddened \hb\ intensity per spaxel in Fig.\,~17n.
No clear trend is seen in contrast to the NGC\,6153 case. We do not find any
obvious trends of the \cpp/\hp\ or \opp/\hp\ ORL abundances (or of the \opp\
ADF) with electron density. No firm trend of the heavy element ORL abundances
with the extinction constant is found in contrast to the case of NGC\,6153. The
detailed parameters of our regression analysis for this nebula are listed in
Table~A1. Parameters of fits obtained when the \opp/\hp\ ORL abundance is based
on the $\lambda$4089 line are also listed in Table A1 (the corresponding plots
are not shown): the fits are in the same direction as previously.

\subsection{NGC~5882}

This PN is the smallest of the three and its symmetrical inner shell was almost
fully captured by the Argus IFU. It also displays lower mean oxygen ADFs and
was meant to be our `control' target. In this section we explore correlations
between the same set of quantities for NGC\,5882 as for the other two nebulae.
The Argus spaxels were grouped based on their distance from the nucleus in the
following manner (cf. Fig.\,~7): points with spaxel coordinates (X, Y) $=$
(4--10,11--18) are coded as red circles whereas those exterior to this boxed
region are coded as blue diamonds.

Fig.\,~18a shows the \opp/\hp\ abundance ratio based on the \foiii\
$\lambda$4959 line {\sl versus} \elt(\foiii) for the 203 spaxels with a
measurement of these quantities (that is, excluding the masked halo region of
the nebula). The solid line is an error-weighted least-squares fit to the data
with a correlation coefficient of $-$0.89. The positive trend between the
oxygen ADF and \elt(\foiii) (Fig.\,~18b; $r$ $=$ 0.52) is not as strong as for
the other two PNe. In Fig.\,~18c--d we plot the \cpp/\hp\ and \opp\/\hp\ ORL
abundance ratios {\sl versus} \elt(\foiii); both show loose positive trends
(with equal $r$ $=$ 0.27). The \opp/\hp\ ORL ratio used in the correlations
plotted in this section is based on the \oii\ $\lambda$4649 line.

The \opp/\hp\ and \cpp\/\hp\ ORL abundances are plotted against the
forbidden-line \opp/\hp\ abundance ratio in Fig.\,~18f--g; loose negative
trends are established in both cases, but in the same vein as for NGC\,6153 and
NGC\,7009 where the trends were stronger. The \cpp/\hp\ ORL abundance is
positively correlated with the \opp/\hp ORL abundance ratio ($r$ $=$ 0.61;
Fig.\,~18e) following the same trend as for the other targets. The \cpp/\opp\
(pure ORL) abundance ratio of NGC\,5882 is approximately constant across the
IFU region with a mean value of 0.52 $\pm$ 0.11.

A loose negative trend is seen between the \opp/\hp\ CEL abundance and the
\hepp\ ionization fraction (mostly for low ionization gas represented by the
blue data points with $r$ $=$ $-$0.68; Fig.\,~18h). The \cpp/\hp ORL abundance
shows a flat trend with increasing \hepp/He (Fig.\,~18i), in contrast to the
cases of NGC\,6153 and 7009, while the \opp\ ADF shows a week positive
correlation with \hepp/He (Fig.\,~18j; $r$ $=$ 0.42). The excitation state of
the gas in this PN is however the lowest of the three, with less than 10 per
cent of helium being doubly ionized (compared with values of up to 20 and 30
per cent for NGC\,6153 and NGC\,7009 respectively).

No clear trends are seen between the doubly ionized carbon and oxygen ORL
abundances, and the \opp\ ADF {\sl versus} the total He/H abundance
(Fig.\,~18k--m) when positive trends were established for the other two
nebulae. Finally, no trend is established between the \opp\ ADF and the \hb\
intensity per spaxel either (Fig.\,~18n). The trend between the \cpp/\hp\ ORL
abundance ratio and the extinction constant is flat and is not shown
(parameters of the fit are however listed in Table~A1). Detailed parameters of
our regression analysis of NGC\,5882 are listed in Table~A1, including those
obtained when the \opp/\hp\ ORL abundance is based on the $\lambda$4089 line;
those fits are in the same vein as previously.

\subsection{Correlations: what do they tell us?}

A strong inverse correlation between the CEL-based \opp/\hp\ abundances and
\elt(\foiii) is established for all three sources. The \opp/\hp\ CEL abundance
decreases by factors of approximately 2.0, 1.5 and 1.3 over $\Delta$\elt's of
$\sim$ $+$1600, $+$1100 and $+$700\,K in NGC\,6153, 7009, and 5882 respectively
(Figs.\,~15d, 17a, 18a). The exponential dependence of CEL-based abundances on
the adopted electron temperature (the Boltzmann factor) mathematically
suppresses most of the scatter in \opp/\hp\ induced by errors in the
\elt(\foiii) determination, thus strengthening the correlation. However, the
correlation mostly reflects a physical process since the variation of
temperature with position is not random (see NGC\,6153, but also NGC\,7009
where regions closer to the nucleus have a higher mean temperature), and
\elt(\foiii) shows a high degree of correlation with other essentially
independent quantities such as the \cpp/\hp\ and \opp/\hp\ ORL abundances. A
physical correlation is expected as well, since \foiii\ is a dominant coolant
of the gas: where the concentration of \opp\ is lower, \elt(\foiii) will tend
to be higher. This may be due to a real variation of the O/H abundance or to a
variation of the \opp/O ionization fraction. Assuming a constant O/H, the
higher forbidden-line temperature in the inner nebular regions can be
qualitatively understood if part of the \opp\ ionic component shifts into
O$^{3+}$; this is a likely possibility since O$^{3+}$ must coexist with \hepp\
(based on ionization stratification considerations), and the \opp/\hp\ CEL
abundance inversely correlates with \hepp/He (Figs.\,~16f, 17h, 18h).

Along the same lines, contrasting NGC\,6153 with NGC\,5882, the observed range
of \elt(\foiii) across the respective IFU regions is larger where the \hepp\
zone is more prominent (i.e., in the former rather than in the latter nebula).
The large opacity of \hep\ and the relatively large average energy of the
corresponding photoelectrons both contribute to further heat the
high-ionization region. That the inner nebula electron temperature can be
explained in this way should await quantitative analysis by means of
photoionization models tailored to these observations. Extra heating can be
provided by the photoelectric effect on small dust grains that may populate the
vicinity of the central stars (see Section~6). The variety of potential heat
sources and the associated uncertainties can make potential spatial variations
of O/H in the gas phase that emits \foiii\ $\lambda$4959 difficult to
ascertain. One intriguing aspect is the fact that \elt(\foiii) continues to
decrease outwards in NGC\,6153, even in regions where \hepp\ and O$^{3+}$
should in principle be minor species. It may be that \hepp\ can survive
relatively farther out from the nucleus than expected as the gas `porosity'
induced by density fluctuations can blur out the theoretical ion
stratification.

The second fundamental piece of information from this analysis is the clear
positive correlation of the \cpp/\hp\ and \opp/\hp\ ORL abundances {\sl versus}
\elt(\foiii) for all three targets -- even for the low ADF `control' NGC\,5882.
Throughout the IFU region of NGC\,6153, for instance, these abundances increase
by a factor of $\sim$\,2 (Fig.\,~15e--f) from the outer to the inner nebula, so
that the corresponding increase in the \opp\ ADF is a factor of $\sim$\,4
(Fig.\,~15c). This is paradoxical in that ORLs should {\sl a priori} be
enhanced in low \elt\ conditions (due to their emissivities obeying an inverse
power law with electron temperature). If a sizeable fraction of the (optical)
\foiii\ flux were to arise from the same parcel of gas emitting a sizeable
fraction of ORL flux, then part of the \foiii\ emission would necessarily be
produced at relatively low electron temperatures and the (observed) average
\elt(\foiii) would be lower (\emph{not higher}), where ORL emission is
strongest. {\it Thus, this paradox demonstrates beyond reasonable doubt that
heavy element CELs and ORLs must essentially arise from unrelated gas
components.}

The `dual abundance' picture of two distinct components of highly ionized gas
at very different temperatures can solve the paradox, but the only known way to
maintain a strongly photoionized gas at low \elt\ is thermostasis by
fine-structure (FS) infrared lines. This mechanism can only be effective for
sufficiently large heavy-element abundances (and sufficiently low electron
densities so that the low critical density FS lines are not quenched). This
interpretation of the data is qualitatively coherent since cool
hydrogen-deficient ionized gas propitiously emits ORLs. Therefore, at another
quantitative level (separate from ORLs), an essential check can be provided by
the 2D spatial distribution of infrared FS line emission; this should be
tractable by means of {\it Spitzer Space Telescope} observations and future
{\it Herschel} surveys.

\section{Discussion and conclusions}

We have presented the first detailed analysis of integral field unit
spectrophotometry of three Galactic PNe (NGC 5882, 6153, and 7009) taken with
the 8.2-m VLT and demonstrated the superb capability of FLAMES Argus data for
the accurate mapping of the physical properties of nebulae. Spatially resolved
maps of 11.5$''$ $\times$ 7.2$''$ areas of the PNe with 0.52$^2$ arcsec$^2$
spaxels were made in the light of \foiii\ $\lambda$4959 and \fneiii\
$\lambda$3967 lines, and the \hb\ $\lambda$4861, \hei\ $\lambda$4471, \heii\
$\lambda$4686, \cii\ $\lambda$4267, \oii\ $\lambda$4089 and $\lambda$4649
recombination lines. The dust extinction and the plasma temperature and density
were mapped out, along with the ionic abundances of helium, \cpp, and \opp,
relative to hydrogen. The \opp/\hp\ ratio was derived from the \oii\ optical
recombination lines and the \foiii\ $\lambda$4959 collisionally excited line.
Maps of the resulting \opp\ abundance discrepancy factor were made. The \opp\
ADF varies between the targets and across the Argus field of view in each case.
In NGC\,6153 and NGC\,7009 the \opp\ ADF and the \cpp\ ORL abundance show
distinct spikes at or near the nucleus. In NGC\,7009 these quantities further
peak at positions coinciding with the high excitation boundary between the
inner nebular shell and the outer envelope; this feature is well registered on
both the Argus maps and the {\it HST} WFPC2 image. In NGC\,5882, where the S/N
ratio of \oii\ ORLs is lower, the \opp ADF derived from the
$\lambda$4089/$\lambda$4959 diagnostics also peaks at the position of the
central star.

In all cases these local maxima are observed over about two spaxels
($\sim$1~arcsec) and are therefore at the limit of our spatial resolution. This
means that the hypothetical structures associated with the hydrogen-poor plasma
should have physical dimensions of $<$ 1000 astronomical units (a.u.), assuming
a typical PN distance of the order of 1\,kpc.

Correlations between the \opp\ ADF and the ionization state of the gas as
gauged by the \hepp/He ionization fraction were established for NGC\,6153 and
7009; likewise strong correlations of the \opp\ ADF and the \cpp/\hp\ and
\opp/\hp\ ORL abundances {\sl versus} the forbidden line electron temperature
were established for these nebulae (for NGC\,5882 these correlations were
weaker but in the same direction). This constitutes new evidence that the
`metallic' ORLs in these PNe are somehow associated with highly ionized plasma,
and that heavy element ORLs and CELs arise from largely unrelated parcels of
gas, in accordance with dual-abundance photoionization models (P\'{e}quignot et
al. 2002). We argue that this hydrogen-deficient, highly ionized plasma may
well (i) have been recently ejected from the nucleus in the form of clumped
gas, or (ii) be part of circumstellar material very close to the star in the
form of a disk-torus structure (based on the spatial correlation of the
diagnostics with the position of the central star), or equally that (iii) it
originates from high excitation gas currently being photoevaporated from
milli-pc scale neutral condensations embedded in the nebulae at the limit of
our spatial resolution.

In case (i), {\it the physical origin of the enhanced heavy-element ORL
emission could be an ensemble of hydrogen-poor clumps analogous to the H-poor
knots observed in the class of nebulae, such as Abell 30 or Abell 58} (e.g.
Borkowski et al. 1993; Guerrero \& Manchado 1996), which may have undergone a
late helium-flash (Iben et al. 1983) or could be associated with neon novae
(Wesson et al. 2007). In this context, the range of ADFs found amongst our
targets could reflect an evolutionary effect with H-poor zones becoming
progressively more readily observed as the density and temperature contrast
between them and the ambient expanding nebula increases over time. One should
recall that NGC\,5882 is positioned in the lower end of the `ADF sequence'
established from the long-slit surveys according to which dense, compact and
lower ionization PNe exhibit smaller integrated abundance discrepancies than
more diffuse, larger and higher ionization nebulae. NGC\,6153 and 7009, on the
other hand, appear to be in a more advanced stage of the `ADF evolution'
(Tsamis et al. 2004; Liu et al. 2004).

In case (ii), {\it the hydrogen-deficient gas could arise from circumstellar
structures} such as those recently discovered by VLT interferometric
observations in association with J-type C-stars (IRAS 18006-3213 -- Deroo et
al. 2007a), post-AGB stars (V390 Vel -- Deroo et al. 2007b) and PNe (Mz\,3 --
Chesneau et al. 2007). These toroidal structures have estimated radii of $<$
1000 a.u. and there is evidence that they contain large amounts of silicate
dust. The inference about their chemistry is important as the C/O ORL abundance
ratio derived for our PN sample is lower than unity; this means that an
association of the hydrogen-poor, O-rich gas with silicate-based (i.e.
oxygen-rich) dust is not unlikely (see below).

In case (iii) {\it the H-poor plasma could originate in evaporating dusty
`cometary knots' analogous to those in the Helix nebula} (NGC\,7293; e.g.
Meaburn et al. 1992) immersed in high ionization nebular zones both near
(NGC\,5882, 6153, 7009) and farther out (NGC\,7009) from the stars. The Helix
globules however are rich in molecular hydrogen (Meixner et al. 2005) and this
hypothesis would require a mechanism for the efficient removal of hydrogen
and/or the chemical differentiation of the evaporated gas for the spectroscopic
signature of H-poor plasma to be observed. Also, in the Helix nebula at least,
there is no evidence of globules inside the high ionization \heii\ zone or of
\heii\ emission from the observed globules, and these features are mostly
associated with emission from lower ionization species (O'Dell et al. 2002;
O'Dell, Henney and Ferland 2007); in contrast, as we have shown, the H-poor
plasma in our PNe is associated with \heii\ emission. On the other hand, this
\emph{does not} preclude that `molecular' globules can be effectively destroyed
in the inner regions of PNe whilst leaving pockets of H-poor plasma behind.
This hypothesis can clearly be put to the test with observations of the type
presented in this work.

We have presented evidence that, in the central regions of NGC\,6153, the \cpp\
ORL abundance, the \hepp/He ionization fraction and the \opp\ ADF are weakly
correlated with the extinction constant. There is marginal evidence that this
may hold for the inner parts of NGC\,7009 as well. This correlation, if real,
could corroborate either the `dusty clump evaporation' scenario or the `dusty
disk' scenario for the origin of the hydrogen-deficient gas. Dust-rich regions
can contribute to the heating of their plasma environs via photoelectric gas
heating (e.g. Dopita \& Sutherland 2000; Stasi{\'n}ska \& Szczerba 2001;
Weingartner, Draine and Barr 2006) and to the metal enrichment of the gas via
grain destruction, thus bringing about this correlation. According to Dopita \&
Sutherland dust grain photoelectric heating is more effective in regions with a
population of small grains, it increases with increasing grain metallicity, and
is more important in highly ionized regions. This effect could also play a role
in the observed rise of \elt(\foiii) in the inner NGC\,6153 regions (in
parallel to usual photoionization which can also enhance the electron
temperature in the \hepp\ zone), although coupled dust/plasma models would be
needed to ascertain and quantify this. Alternatively, since the theoretical
H$\gamma$/\hb\ line ratio decreases with decreasing electron temperature
(Storey \& Hummer 1995), and since the extinction constant (based on
H$\gamma$/\hb) was computed here adopting \elt(\foiii), the positive
correlation of $c$(\hb) with the ADF may be indicative of the fact that \hi\
lines in PNe partly arise from cool gas which emits most of the heavy element
ORLs, in accordance with dual-abundance models.

Relatively weak correlations between the \opp\ ADF and the \opp/\hp, \cpp/\hp\
ORL abundances {\sl versus} the He/H ratio in NGC\,6153 and (in the inner
regions) of NGC\,7009 have been established. This result, which should be
backed up by further studies, constitutes new observational evidence that
corroborates the predictions of the dual-abundance PN models of P\'{e}quignot
et al. (2002, 2003). In those models, the heavy element-rich (H-poor) gas
component, which is responsible for the bulk of the heavy element ORL emission,
is also helium-rich and as such it contributes non-negligibly to the emissivity
of the helium ORLs. When this effect is not accounted for, as in empirical
methods like the one we have used in this work, there is an insidious bias
towards overestimating the helium to hydrogen abundance ratio. Based on our
results it is therefore clear that total He/H abundances of PNe as
traditionally determined from empirical methods, and which assume chemical
uniformity of the plasma, will be systematically overestimated. Again detailed
modelling in the context of a chemically inhomogeneous nebula is needed to
remove this bias. With respect to helium it should be noted that, in contrast
to \hepp, \hep\ is not perturbed by H-poor material, since nebular regions with
high \hepp/\hp\ give the same (\hep\ $+$ \hepp)/H ratio as regions with low
\hep/\hp.

In conclusion, our detailed IFS analysis of a few suitable objects has allowed
us to identify important new elements of the still incomplete puzzle of
planetary nebula astrophysics, and we have put forward testable proposals for
the likely nature of some of the missing pieces. Furthermore, this study has
underlined a number of clear requirements that a detailed solution to the `ORL
vs. CEL' problem should meet, if it were to be successful.

\vspace{7mm} \noindent {\bf Acknowledgments}

We thank A. Blecha and G. Simond of the Geneva Observatory for answering our
queries on girBLDRS and for providing extra calibrations and patches. Marina
Rejkuba is thanked for early advice on the reduction of Argus data. Francesca
Primas is thanked for an informed explanation of the `scattered light' issue
mentioned in Section~3. We appreciate comments from Bob O'Dell and thank the
referee for a constructive report. This study has made use of the NASA ADS
database. YGT gratefully acknowledges the hospitality of ESO Garching where
part of this work was completed during a science visit.

\appendix

\section{}

\begin{table*}
\caption{Parameters of least-squares regression analysis discussed in
Section~5: A -- intercept, B -- slope, $r$ -- correlation coefficient, $\sigma$
-- standard deviation, Points -- number of data points fitted. Values in
parentheses are exponents of base 10.}
\begin{tabular}{lccrll}
\noalign{\vskip3pt} \noalign{\hrule} \noalign{\vskip3pt}

Diagnostics                      &A                    &B                      &$r$          &$\sigma$   &Points \\
\noalign{\vskip3pt} \noalign{\hrule}\noalign{\vskip3pt}

\noalign{\vskip3pt}
&\multicolumn{5}{c}{NGC~6153}                                        \\
\opp/\hp\ {\sc cel} vs. \elt(\foiii)  &2.25($-$3) $\pm$ 2.00($-$5)   &$-$1.99($-$7) $\pm$ 2.20($-$9) &$-$0.98      &0.57       &296    \\
ADF(\opp $\lambda$4649) vs. \elt(\foiii)      &$-$58.2 $\pm$ 1.70     &7.39($-$3) $\pm$ 1.90($-$4)     &0.92       &0.82       &296     \\
ADF(\opp $\lambda$4089) vs. \elt(\foiii)      &$-$96.7 $\pm$ 3.70     &0.012 $\pm$ 4.11($-$4)     &0.86       &2.2       &295$^a$     \\
\cpp/\hp\ {\sc orl} vs. \elt(\foiii)  &$-$9.56($-$3) $\pm$ 3.79($-$4)     &1.33($-$6) $\pm$ 4.19($-$8)     &0.88       &2.2     &295     \\
\opp/\hp\ $\lambda$4649 {\sc orl} vs. \elt(\foiii)  &$-$9.13($-$3) $\pm$ 7.66($-$4)     &1.42($-$6) $\pm$ 8.40($-$8)     &0.70       &1.2     &296     \\

\opp/\hp\ $\lambda$4089 {\sc orl} vs. \elt(\foiii)  &$-$2.34($-$2) $\pm$ 1.56($-$3)     &3.15($-$6) $\pm$ 1.72($-$7)     &0.73       &2.8     &291$^a$     \\
\cpp/\hp\ {\sc orl} vs. \opp/\hp\ $\lambda$4649 {\sc orl} &6.82($-$5) $\pm$ 8.13($-$5)     &0.645 $\pm$ 0.021     &0.87       &2.3     &294    \\
\cpp/\hp\ {\sc orl} vs. \opp/\hp\ $\lambda$4089 {\sc orl} &6.47($-$4) $\pm$ 6.86($-$5)     &0.347 $\pm$ 0.013     &0.85       &2.4     &290$^a$    \\

\opp/\hp\ $\lambda$4649 {\sc orl} vs. \opp/\hp\ {\sc cel} &6.72($-$3) $\pm$ 1.80($-$4)     &$-$6.46 $\pm$ 0.39     &$-$0.69       &1.3     &296    \\

\opp/\hp\ $\lambda$4089 {\sc orl} vs. \opp/\hp\ {\sc cel} &1.15($-$2) $\pm$ 3.81($-$4)     &$-$14.1 $\pm$ 0.82     &$-$0.71       &2.8     &291$^a$    \\

\cpp/\hp\ {\sc orl} vs. \opp/\hp\ {\sc cel} &5.16($-$3) $\pm$ 9.72($-$5)     &$-$5.87 $\pm$ 0.20     &$-$0.86       &2.4     &295    \\
\opp/\hp\ {\sc cel} vs. \hepp/He      &5.45($-$4) $\pm$ 5.77($-$6)     &$-$1.36($-$3) $\pm$ 4.48($-$5)     &$-$0.87       &1.5     &296    \\
\cpp/\hp\ {\sc orl} vs. \hepp/He      &1.72($-$3) $\pm$ 3.35($-$5)     &9.83($-$3) $\pm$ 3.58($-$4)     &0.85       &2.5     &295    \\
ADF(\opp $\lambda$4649) vs. \hepp/He          &4.62 $\pm$ 0.15     &53.1 $\pm$ 1.90     &0.86       &1.1     &295    \\
ADF(\opp $\lambda$4089) vs. \hepp/He          &5.17 $\pm$ 0.24     &87.4 $\pm$ 3.20     &0.85       &2.3     &293$^a$    \\
\opp/\hp\ $\lambda$4649 {\sc orl} vs. He/H          &$-$0.0127 $\pm$ 0.0015     &0.125 $\pm$ 0.011     &0.55       &1.5     &296    \\
\opp/\hp\ $\lambda$4089 {\sc orl} vs. He/H          &$-$0.0321 $\pm$ 0.0031     &0.282 $\pm$ 0.024     &0.57       &3.3     &289$^a$    \\
\cpp/\hp\ {\sc orl} vs. He/H          &$-$0.0118 $\pm$ 0.0010      &0.109 $\pm$ 0.007     &0.68       &3.4     &295   \\
ADF(\opp $\lambda$4649) vs. He/H              &$-$54.2 $\pm$ 4.90     &476 $\pm$ 37.9     &0.59       &1.6     &296   \\
ADF(\opp $\lambda$4089) vs. He/H              &$-$105 $\pm$ 9.90     &882 $\pm$ 75.8     &0.57       &3.6     &289$^a$   \\
\cpp/\hp\ {\sc orl} vs. $c$(\hb)          &$-$5.50($-$4) $\pm$ 4.57($-$4)     &3.03($-$3) $\pm$ 3.94($-$4)    &0.54       &1.9     &148$^a$   \\

ADF(\opp $\lambda$4649) vs. $c$(\hb)          &$-$8.86 $\pm$ 3.28     &18.0 $\pm$ 2.84   &0.47       &1.0     &147$^{a, b}$   \\
ADF(\opp $\lambda$4089) vs. $c$(\hb)          &$-$19.4 $\pm$ 4.55     &31.8 $\pm$ 3.90   &0.55       &1.9     &147$^{a, b}$   \\
\elt(\foiii) vs. $c$(\hb)                     &7530 $\pm$ 248     &1693 $\pm$ 211   &0.56       &1.2     &146$^{a, b}$   \\

\hepp/He vs. $c$(\hb)          &$-$0.167 $\pm$ 0.026     &0.253 $\pm$ 0.023   &0.68       &2.0     &147$^b$   \\
\hep/\hp\ vs. \hepp/\hp\        &0.128 $\pm$ 3.41($-$4)     &$-$0.673 $\pm$ 0.022     &$-$0.87       &0.45     &292$^{a, b}$   \\

\noalign{\vskip3pt}
&\multicolumn{5}{c}{NGC~7009}                                        \\
\noalign{\vskip3pt}
\opp/\hp\ {\sc cel} vs. \elt(\foiii)  &1.63($-$3) $\pm$ 1.52($-$5)   &$-$1.23($-$7) $\pm$ 1.47($-$9) &$-$0.98      &3.1       &294    \\
ADF(\opp $\lambda$4649) vs. \elt(\foiii)      &$-$26.2 $\pm$ 1.40   &3.05($-$3) $\pm$ 1.41($-$4) &0.78      &2.5       &296    \\
ADF(\opp $\lambda$4089) vs. \elt(\foiii)      &$-$50.8 $\pm$ 2.40   &5.78($-$3) $\pm$ 2.34($-$4) &0.82      &2.0       &293$^a$    \\
\cpp/\hp\ {\sc orl} vs. \elt(\foiii)  &$-$3.23($-$3) $\pm$ 4.23($-$4)   &4.08($-$7) $\pm$ 4.06($-$8) &0.66      &1.6       &135$^b$    \\
\opp/\hp\ $\lambda$4649 {\sc orl} vs. \elt(\foiii)  &$-$9.48($-$3) $\pm$ 9.88($-$4)   &1.08($-$6) $\pm$ 9.48($-$8) &0.70      &1.6       &135$^b$    \\
\opp/\hp\ $\lambda$4089 {\sc orl} vs. \elt(\foiii)  &$-$8.56($-$3) $\pm$ 8.72($-$4)   &1.13($-$6) $\pm$ 8.49($-$8) &0.62      &2.0       &293$^a$    \\
\cpp/\hp\ {\sc orl} vs. \opp/\hp\ $\lambda$4649 {\sc orl} &7.50($-$5) $\pm$ 3.23($-$5)   &0.519 $\pm$ 0.018 &0.86      &1.8       &296    \\
\cpp/\hp\ {\sc orl} vs. \opp/\hp\ $\lambda$4089 {\sc orl} &3.11($-$4) $\pm$ 3.31($-$5)   &0.230 $\pm$ 0.011 &0.78      &2.2       &293$^a$    \\
\opp/\hp\ $\lambda$4649 {\sc orl} vs. \opp/\hp\ {\sc cel} &4.78($-$3) $\pm$ 2.81($-$4)   &$-$8.70 $\pm$ 0.811 &$-$0.68      &1.6       &135$^b$    \\
\opp/\hp\ $\lambda$4089 {\sc orl} vs. \opp/\hp\ {\sc cel} &6.10($-$3) $\pm$ 2.53($-$4)   &$-$8.41 $\pm$ 0.694 &$-$0.58      &2.0       &293$^a$    \\%
\cpp/\hp\ {\sc orl} vs. \opp/\hp\ {\sc cel} &2.15($-$3) $\pm$ 1.22($-$4)   &$-$3.27 $\pm$ 0.350 &$-$0.63      &1.6       &135$^b$    \\
\opp/\hp\ {\sc cel} vs. \hepp/He      &4.39($-$4) $\pm$ 2.90($-$6)   &$-$4.22($-$4) $\pm$ 1.64($-$5) &$-$0.84      &7.9       &289    \\
\cpp/\hp\ {\sc orl} vs. \hepp/He      &7.39($-$4) $\pm$ 6.37($-$5)   &1.36($-$3) $\pm$ 3.10($-$4) &0.35      &1.9       &135$^b$    \\
ADF(\opp $\lambda$4649) vs. \hepp/He          &1.19 $\pm$ 0.53   &19.3 $\pm$ 2.57 &0.55      &2.6       &135$^b$    \\
ADF(\opp $\lambda$4089) vs. \hepp/He          &5.78 $\pm$ 0.28   &15.9 $\pm$ 1.62 &0.50      &3.0       &293$^a$    \\
\opp/\hp\ $\lambda$4649 {\sc orl} vs. He/H          &$-$4.79($-$4) $\pm$ 4.64($-$4)   &0.0201 $\pm$ 0.0040 &0.28      &2.5       &294    \\
\opp/\hp\ $\lambda$4089 {\sc orl} vs. He/H          &$-$6.38($-$3) $\pm$ 1.11($-$3)   &0.0819 $\pm$ 0.0096 &0.45      &2.3       &290$^a$    \\
\cpp/\hp\ {\sc orl} vs. He/H          &$-$3.13($-$4) $\pm$ 2.88($-$4)   &0.0115 $\pm$ 0.0025 &0.26      &3.4       &295    \\
ADF(\opp $\lambda$4649) vs. He/H              &$-$19.5 $\pm$ 2.01    &214 $\pm$ 17.5  &0.58      &3.3       &295    \\
ADF(\opp $\lambda$4089) vs. He/H              &$-$38.5 $\pm$ 3.90    &409 $\pm$ 34.0  &0.58      &2.9       &293$^a$    \\
\hep/\hp\ vs. \hepp/\hp\        &0.110 $\pm$ 3.64($-$4)     &$-$0.789 $\pm$ 0.018     &$-$0.93       &0.86     &293$^a$   \\

\noalign{\vskip3pt}
&\multicolumn{5}{c}{NGC~5882}                                        \\
\noalign{\vskip3pt}
\opp/\hp\ {\sc cel} vs. \elt(\foiii)  &1.71($-$3) $\pm$ 4.52($-$5)   &$-$1.34($-$7) $\pm$ 4.72($-$9) &$-$0.89      &8.2       &204    \\
ADF(\opp $\lambda$4649) vs. \elt(\foiii)      &$-$5.39 $\pm$ 0.87    &7.92($-$4) $\pm$ 9.09($-$5) &0.52      &1.0       &202    \\
ADF(\opp $\lambda$4089) vs. \elt(\foiii)      &$-$21.8 $\pm$ 2.62    &2.70($-$3) $\pm$ 2.74($-$4) &0.58      &1.8       &197$^a$    \\
\cpp/\hp\ {\sc orl} vs. \elt(\foiii)  &$-$3.55($-$4) $\pm$ 2.06($-$4)   &8.75($-$8) $\pm$ 2.16($-$8) &0.27      &1.8       &203    \\
\opp/\hp\ $\lambda$4649 {\sc orl} vs. \elt(\foiii)  &$-$6.16($-$4) $\pm$ 3.85($-$4)   &1.60($-$7) $\pm$ 4.05($-$8) &0.27      &1($-$4)       &203    \\
\opp/\hp\ $\lambda$4089 {\sc orl} vs. \elt(\foiii)  &$-$2.21($-$3) $\pm$ 1.13($-$3)   &4.07($-$7) $\pm$ 1.19($-$8) &0.24      &1.8       &190$^a$    \\
\cpp/\hp\ {\sc orl} vs. \opp/\hp\ $\lambda$4649 {\sc orl} &1.56($-$4) $\pm$ 2.98($-$5)   &0.347 $\pm$ 0.031 &0.61      &1.5       &203    \\
\cpp/\hp\ {\sc orl} vs. \opp/\hp\ $\lambda$4089 {\sc orl} &5.13($-$4) $\pm$ 1.90($-$5)   &$-$0.0164 $\pm$ 0.0109 &$-$0.11      &1.9       &196$^a$    \\
\opp/\hp\ $\lambda$4649 {\sc orl} vs. \opp/\hp\ {\sc cel} &1.37($-$3) $\pm$ 1.12($-$4)   &$-$1.07 $\pm$ 0.260 &$-$0.28      &1($-$4)       &203    \\
\opp/\hp\ $\lambda$4089 {\sc orl} vs. \opp/\hp\ {\sc cel} &2.95($-$3) $\pm$ 3.15($-$4)   &$-$3.01 $\pm$ 0.739 &$-$0.28      &1.82       &193$^a$    \\
\cpp/\hp\ {\sc orl} vs. \opp/\hp\ {\sc cel} &7.13($-$4) $\pm$ 5.95($-$5)   &$-$0.547 $\pm$ 0.140  &$-$0.26      &1.8       &203    \\

\noalign{\vskip3pt} \noalign{\hrule}\noalign{\vskip3pt}
\end{tabular}
\end{table*}

\begin{table*}
\caption{{\it --continued}}
\begin{tabular}{lccrll}
\noalign{\vskip3pt} \noalign{\hrule} \noalign{\vskip3pt}
Diagnostics                      &A                    &B                      &$r$          &$\sigma$   &Points \\
\noalign{\vskip3pt} \noalign{\hrule}\noalign{\vskip3pt}
\noalign{\vskip3pt}

\opp/\hp\ {\sc cel} vs. \hepp/He      &4.47($-$4) $\pm$ 2.97($-$6)   &$-$8.48($-$4) $\pm$ 7.62($-$5) &$-$0.68      &14       &148$^c$    \\
\cpp/\hp\ {\sc orl} vs. \hepp/He      &4.88($-$4) $\pm$ 7.70($-$6)   &$-$1.44($-$4) $\pm$ 1.73($-$4) &$-$0.06      & 1.9       &199    \\
ADF(\opp $\lambda$4649) vs. \hepp/He          &1.95 $\pm$ 0.04   &5.62 $\pm$ 0.87 &0.42      &  0.31       &200    \\
ADF(\opp $\lambda$4089) vs. \hepp/He          &3.52 $\pm$ 0.13   &12.6 $\pm$ 3.19 &0.27      &  2.7       &200$^a$    \\
\opp/\hp\ $\lambda$4649 {\sc orl} vs. He/H          &2.23($-$3) $\pm$ 2.94($-$4)   &$-$0.012 $\pm$ 0.003 &$-$0.30      &1($-$4)        &203    \\
\opp/\hp\ $\lambda$4089 {\sc orl} vs. He/H          &4.87($-$3) $\pm$ 1.22($-$3)   &$-$0.029 $\pm$ 0.011 &$-$0.18      &2.6        &203$^a$    \\
\cpp/\hp\ {\sc orl} vs. He/H          &9.29($-$4) $\pm$ 1.72($-$4)   &$-$4.12($-$3) $\pm$ 1.59($-$3) &$-$0.18      &1.9        &203    \\
\cpp/\hp\ vs. $c$(\hb)          &4.76($-$4) $\pm$ 8.1($-$6)     &1.38($-$5) $\pm$ 1.61($-$5)    &0.06       &1.9     &203$^a$   \\
ADF(\opp $\lambda$4649) vs. He/H              &7.07 $\pm$ 0.84     &$-$45.4 $\pm$ 7.80 &$-$0.38      &0.31        &203    \\
ADF(\opp $\lambda$4089) vs. He/H              &12.9 $\pm$ 2.46     &$-$83.1 $\pm$ 22.7 &$-$0.25      &2.2        &197$^a$    \\
\hep/\hp\ vs. \hepp/\hp\        &0.108 $\pm$ 3.27($-$4)     &$-$1.13 $\pm$ 0.071     &$-$0.75       &1.4     &201$^a$   \\

\noalign{\vskip3pt} \noalign{\hrule}\noalign{\vskip3pt}
\end{tabular}
\begin{description}
\item[$^a$] Corresponding plot not shown.  \item[$^b$] Fit only for the inner nebula (coloured red) data points; see text
for details. \item[$^c$] Fit only for the outer nebula (coloured blue) data points; see text for details.

\end{description}
\end{table*}


\end{document}